\documentclass[a4paper,10pt]{article}
\usepackage{graphicx}
\usepackage{comment}
\usepackage{amssymb}
\usepackage{amsmath}
\usepackage{nicefrac}
\usepackage{dcolumn}
\usepackage{bm}

\begin{document}

\huge

\begin{center}
DETAILED OPACITY CALCULATIONS FOR ASTROPHYSICAL APPLICATIONS
\end{center}

\vspace{0.5cm}

\large

\begin{center}
Jean-Christophe Pain\footnote{jean-christophe.pain@cea.fr}, Franck Gilleron and Maxime Comet
\end{center}

\normalsize

\begin{center}
\it CEA, DAM, DIF, F-91297 Arpajon, France
\end{center}

\vspace{0.5cm}

\begin{abstract}
Nowadays, several opacity codes are able to provide data for stellar structure models, but the computed opacities may show significant differences. In this work, we present state-of-the-art precise spectral opacity calculations, illustrated by stellar applications. The essential role of laboratory experiments to check the quality of the computed data is underlined. We review some X-ray and XUV laser and Z-pinch photo-absorption measurements as well as X-ray emission spectroscopy experiments involving hot dense plasmas produced by ultra-high-intensity laser irradiation. The measured spectra are systematically compared with the fine-structure opacity code SCO-RCG. Focus is put on iron, due to its crucial role in understanding asteroseismic observations of $\beta$ Cephei-type and Slowly Pulsating B stars, as well as of the Sun. For instance, in $\beta$ Cephei-type stars, the iron-group opacity peak excites acoustic modes through the ``kappa-mechanism''. A particular attention is paid to the higher-than-predicted iron opacity measured at the Sandia Z-machine at solar interior conditions. We discuss some theoretical aspects such as density effects, photo-ionization, autoionization or the ``filling-the-gap'' effect of highly excited states.
\end{abstract}

\section{Introduction}\label{int}

The proper accounting for spectral lines is a keypoint of hot-plasma modelling, for radiative transfer (mostly through Rosseland mean opacities), and more generally for absorption and emission spectroscopy, where they are used as a diagnostic tool to infer the temperature (from ratios of line intensities) and density (from line widths) of plasmas. The applications range from inertial confinement fusion, magnetic confinement fusion (for the radiative losses in the W-coated tiles of divertor) and astrophysics. Heating by fusion reactions deep within stellar cores produces thermal X-ray radiation and models of stellar structure and evolution are very sensitive to radiative transfer and opacity. Matter at extreme densities and temperatures gives us a chance to study important and exotic physical processes: plasmon neutrinos as a test of electro-weak theory at low-energies, turbulent energy transport in high-gravity environments, dark matter in the form of axions \cite{schlattl99} and/or WIMPS, \emph{etc.} \cite{vinyoles15}. Exploration of extreme physics in star interiors is possible thanks to asteroseismology. The Cepheid stars (named after $\delta$ Cephei) are known to be used to estimate the stellar distances. They have been better understood after a revision by a factor two of the opacity for species of atomic number $Z>2$ \cite{simon82} confirmed by the first measurements of absorption coefficients \cite{dasilva92}. The $\beta$ Cephei-type stars (which should not be confused with Cepheid variables), also known as $\beta$ Cepheids and named after $\beta$ Cephei (or Alfirk) are now of primary interest due to the correlation between the frequencies of the excited modes and the mass of these stars for a given age. Such stars are now used to estimate the age of several young open clusters \cite{balona97}. In $\beta$ Cephei-type stars, the iron-group (chromium, iron and nickel) opacity peak (when plotting Rosseland opacity with respect to temperature) excites acoustic modes through the ``kappa-mechanism'' \cite{eddington26a,eddington26b}. 

With the recent revision of solar chemical abundances \cite{asplund09}, the standard stellar model fails to reproduce the helioseismic results. Among the possible explanations, an increase of 5 to 20 \% of the opacity in the solar radiative zone would be sufficient to reconcile the modelling and the asteroseismic observations. Several absorption spectroscopy measurements were performed recently on the Z machine at Sandia National Laboratory by Bailey {\emph et al.} \cite{bailey09} at conditions similar to those at the boundary between the convective and radiative zones of the Sun. In that region, iron is responsible for 25 \% of the total opacity. One of those experiments, published in 2015 \cite{bailey15} revealed that, in the spectral range between 7 and 12.7 \AA, the opacity inferred from the measurements was higher than the opacity predicted by all the best codes in the world by 30 to 400 \%.

Highly charged iron produces some of the brightest X-ray emission lines from hot astrophysical objects, including galaxy clusters and stellar coronae, and dominates the emission of the Sun at wavelengths near 15 \AA. The Fe XVII spectrum is, however, poorly fitted by the best astrophysical models. A particular problem has been that the strongest Fe XVII line is weaker than predicted. This has affected the interpretation of observations by the Chandra and XMM-Newton orbiting X-ray missions \cite{paerels03}, fuelling a continuing controversy over whether this discrepancy is caused by an incomplete modelling of the plasma environment in these objects or by shortcomings in the treatment of the underlying atomic physics. 

Of course, iron is not the only element we are interested in. To explore the history of star formation in our galaxy, cosmochronology estimates the ages of the coolest white dwarf stars. Measuring relative line shapes of H$_{\beta}$, H$_{\gamma}$, and H$_{\delta}$ enables one to determine white dwarf photospheric and atmospheric conditions \cite{wiese72}. 

In Sec. \ref{rad}, the different processes contributing to radiative opacity are introduced, and the detailed opacity code SCO-RCG is presented. Throughout the last years, the code was used to interpret several laser or Z-pinch experiments, as shown in Sec. \ref{exp}. In Sec. \ref{env}, we discuss the role of opacities for the modelling of stellar envelopes, and present the kappa mechanism, responsible for the pulsations of particular stars. In Sec. \ref{bou}, the importance of the opacity at the boundary between convective and radiative zones of the Sun (also named ``tachocline'' \cite{spiegel92}) is outlined, and theoretical efforts dedicated to understanding the experiment performed recently on the Z machine at Sandia National Laboratory are detailed in Sec. \ref{bai}. The enigma of the ratio between two important iron lines, characteristic of the corona of the Sun and of other stars such as $\alpha$ Aurigae (Capella) or $\beta$ Canis Majoris (Procyon), is briefly mentioned in Sec. \ref{3c3}, and the need for Stark-broadening modelling of hydrogen Balmer lines is recalled in Sec. \ref{whi}.

\subsection{Radiative opacity}\label{rad}

The spectral opacity is the photo-absorption cross-section per mass unit, usually expressed in cm$^2$/g. The total frequency-dependent opacity can be calculated as the sum of the contributions of different processes: photo-excitation $\kappa_{\mathrm{bb}}$, photo-ionization $\kappa_{\mathrm{bf}}$, inverse Bremsstrahlung $\kappa_{\mathrm{ff}}$ and photon scattering $\kappa_{\mathrm{scat}}$. It is then given by the following expression \cite{mayer47}:

\begin{equation}
\kappa'(h\nu)=\kappa(h\nu)\left(1-e^{-\left(h\nu/k_BT\right)}\right)+\kappa_{\mathrm{scat}}(h\nu),
\end{equation}

\noindent where $h$ the Planck constant, $k_B$ the Boltzmann constant, $T$ the temperature, $\nu$ the photon frequency and

\begin{equation}
\kappa(h\nu)=\kappa_{\mathrm{bb}}(h\nu)+\kappa_{\mathrm{bf}}(h\nu)+\kappa_{\mathrm{ff}}(h\nu).
\end{equation}

\noindent Photo-excitation and de-excitation can be described as

\begin{equation}
X_i^{q+}+h\nu \leftrightharpoons X_j^{q+},
\end{equation}

\noindent where $X_i^{q+}$ is an ion with charge $q$ in an excitation state $i$. The signature of the emitted or absorbed photon $h\nu$ is a spectral line. The relevant atomic parameter for the direct and inverse processes is the oscillator strength ($f$), and one has \cite{pain17b}

\begin{equation}
\kappa_{\mathrm{bb}}(h\nu)=\frac{1}{4\pi\epsilon_0}\frac{\mathcal{N}_A}{A}\frac{\pi e^2h}{mc}\sum_{i\rightarrow j}\mathcal{P}_if_{i\rightarrow j}\Psi_{i\rightarrow j}(h\nu),
\end{equation}

\noindent where $\mathcal{P}_i$ is the population of initial level $i$, $f_{i\rightarrow j}$ the oscillator strength and $\Psi_{i\rightarrow j}$ the profile of the spectral line corresponding to the transition $i\rightarrow j$, accounting for broadening mechanisms (Doppler, Stark, ...). $\epsilon_0$ is the dielectric constant, $\mathcal{N}_A$ the Avogadro number, $e$ and $m$ represent respectively the electron charge and mass, $c$ the speed of light and $A$ the atomic mass of the considered element. Photo-ionization is a process that occurs when a bound electron $e^-$ is ejected after absorption of a photon

\begin{equation}
X_i^{q+}+h\nu \leftrightharpoons X_j^{(q+1)+}+e^-.
\end{equation}

\noindent The inverse process is radiative recombination (RR), occuring when a free electron recombines with an ion along with emission of a photon. This can occur via an intermediate excited state as

\begin{equation}
e^-+X_i^{q+}\leftrightharpoons \left(X_j^{(q-1)+*}\right)\leftrightharpoons \left\{\begin{array}{ll}
e^-+X_k^{q+} & AI\\
X_m^{(q-1)+}+h\nu & DR
\end{array}\right..
\end{equation}

\noindent A colliding electron excites the target and attaches to form the short-lived autoionizing state (*), which may be doubly excited, but not necessarily: it is only required for it to lie above the first ionization limit, and to be able to decay radiatively to bound states. The intermediate state decays either by autoionization (AI) where the electron becomes free and the target drops to ground state, or by dielectronic recombination (DR) where the electron gets bound by emitting a photon. Photo-ionization resonances can be seen in absorption spectra and DR resonances in emission spectra. The opacity involves also two other processes, inverse Bremsstrahlung or free-free absorption, and photon-electron scattering. Bremsstrahlung refers to the radiation emitted by an electron slowing down in the electromagnetic field of an ion. The inverse process occurs when a free electron and an ion absorb a photon 

\begin{equation}
h\nu+\left[X_i^{q+}+e^-(\epsilon)\right]\rightarrow X_j^{q+}+e^-\left(\epsilon'\right),
\end{equation}

\noindent $\epsilon$ and $\epsilon'$ being the energies of the free electron before and after the photo-absorption. Calculations of the free-free cross-section involve quantities related to the elastic-scattering matrix elements for electron-impact excitation of ions. The detailed (fine-structure) opacity code SCO-RCG \cite{porcherot11,pain15a,pain15b} enables one to compute precise opacities for the calculation of accurate Rosseland means (see Sec. \ref{ross}). The (super-)configurations are generated by the SCO code \cite{blenski00} on the basis of a statistical fluctuation theory (see the details below) and a self-consistent computation of atomic structure is performed for all the configurations. In such a way, each configuration has its own set of wavefunctions. The latests are determined in a single-configuration approximation. One peculiarity of the code is that it does not rely on the ``isolated atom'' picture, but on a realistic atom-in-plasma modelling (equation of state). Relativistic effects are taken into account in the Pauli approximation. The DLA (Detailed Line Accounting) part of the spectrum is performed using an adapted version of the RCG routine from Cowan's suite of atomic-structure and spectra codes \cite{cowan81}. The RCG source code was used for decades by spectroscopists, it has many available options and is well documented. In SCO-RCG, criteria are defined to select transition arrays that can be treated line-by-line. The data required for the calculation of the detailed transition arrays (Slater, spin-orbit and dipolar integrals) are obtained from SCO, providing in this way a consistent description of the plasma screening effects on the wavefunctions. Then, the level energies and the lines are calculated by RCG. The computation starts with an average-atom calculation in local thermodynamic equilibrium (LTE), which provides the average populations of the subshells. We then build a list of super-configurations of the kind

\begin{equation}\label{sc}
(1s)^{p_1}(2s)^{p_2}...\left(n_{k-1}\ell_{k-1}\right)^{p_{k-1}}\left(\prod_{i=k}^Nn_i\ell_i\right)^{p_k},
\end{equation}

\noindent each super-configuration containing $k$ supershells and $N$ subshells, where $n_N\ell_N$ is the last highest-energy subshell found by the average-atom calculation at the given density and temperature. A super-configuration is made of supershells (a supershell being a group of subshells) populated by electrons. In Eq. (\ref{sc}), the first $(k-1)$ supershells are ordinary subshells and the last $(N-k+1)$ subshells are gathered in the same supershell. We use the LTE fluctuation theory \cite{perrot00} around the average-atom non-integer populations in order to determine the range of variation of the populations $p_k$, $k$=1, $N$ and therefore the possible list of configurations (if $p_N$ is equal to zero) or super-configurations (if $p_N$ is strictly positive). The super-configurations are then sorted according to their Boltzmann weights, estimated using the average-atom wavefunctions. We only keep the $N$ (super-)configurations having the highests weights. The strength of this approach is that it enables one to take into account many highly excited states and satellite lines. The populations of those states may be small, but their number is so huge that they can play a significant role in the opacity. The orbitals in the Rydberg supershell are chosen in a way so that they weakly interact with inner orbitals. A DLA calculation is performed if possible and necessary for all the transition arrays starting from the configuration; DLA computations are carried out only for pairs of configurations giving rise to less than 800,000 lines. In other cases, transition arrays are represented by Gaussian profiles in the UTA (Unresolved Transition Array \cite{bauche79,bauche82}) or SOSA (Spin-Orbit Split Array \cite{bauche85}) formalisms. If the Rydberg supershell contains at least one electron, then transitions starting from the super-configuration are taken into account by the Super Transition Array (STA) model \cite{bar89}. The amount of detailed calculations performed in SCO-RCG is now largely dominant.

The PRTA (Partially Resolved Transition Array) model was recently implemented. It enables one to replace many statistical transition arrays by small-scale DLA calculations. We have extended this approach to the STA formalism \cite{bar89}, consisting in omitting the Rydberg supershell in the computation, and adding its contribution to the widths of all lines. The contribution of the Rydberg supershell is included as a Gaussian ``dressing function'' \cite{pain15a}. We have also the possibility to replace this dressing function by a coarse grained configurationally resolved profile, following the CRSTA (Configurationally Resolved Transition Array) method \cite{kurzweil16}. 

\section{Interpretation of experiments}\label{exp}

\begin{figure}[h]
\vspace{3mm}
\centering
\includegraphics[width=7.2cm]{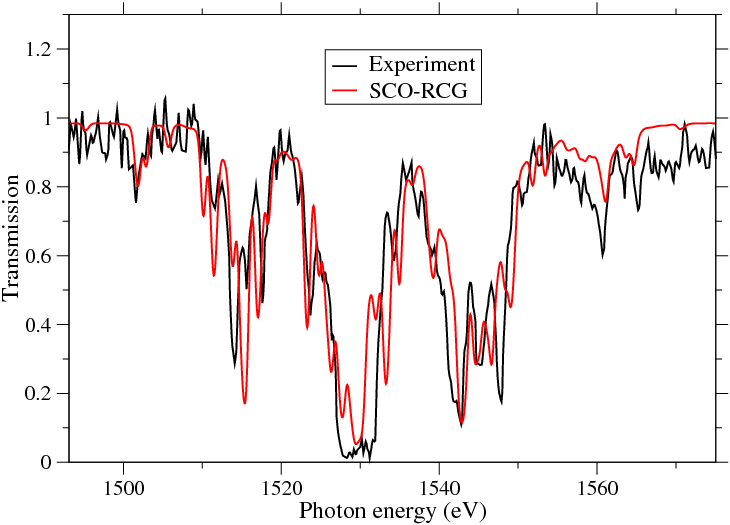}
\hspace{2mm}
\includegraphics[width=7cm]{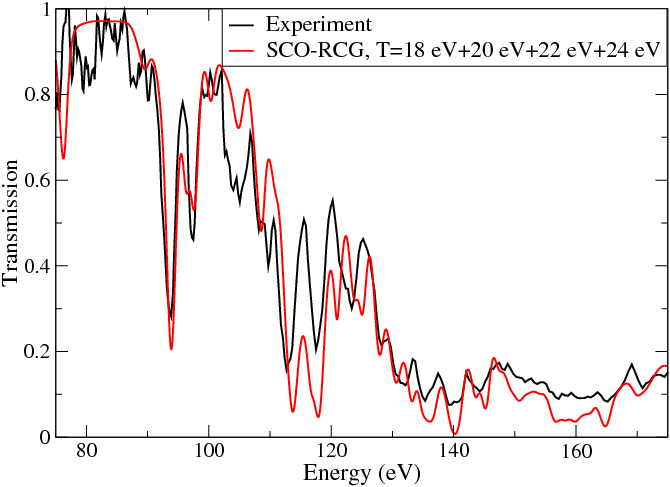}
\caption{Left: aluminum transmission spectrum measured by Davidson \emph{et al.} \cite{davidson88} on HELEN laser facility in the United Kingdom and interpreted by SCO-RCG at $T$=37 eV and $\rho$=0.01 g/cm$^3$. The areal mass is equal to 54 $\mu$g/cm$^{2}$. Right: aluminum spectrum measured by Winhart \emph{et al.} \cite{winhart95,winhart96} on ASTERIX IV laser facility in Germany and interpreted by SCO-RCG at $\rho$=0.01 g/cm$^3$ and averaged over four temperatures ($T$=18, 20, 22 and 24 eV) in order to simulate the gradients. The areal mass is equal to 30 $\mu$g/cm$^{2}$.}
\label{Al_Eidmann2+davidson3}
\end{figure}

In the experiment by Davidson \emph{et al.} \cite{davidson88}, the laser beams were produced by a neodymium-doped glass laser system, HELEN, delivering about 100 J of 1.06 $\mu$m radiation in 300 ps, which is converted to about 60 J of 0.53 $\mu$m light incident on a target, in each of its two beams. One beam is used to directly irradiate a thin layer of gold (500 \AA) producing X-rays that serve as an heating source for an aluminum sample (0.2-0.5 $\mu m$). A layer of plastic is included (0.25 $\mu m$), which is largely transparent to these X rays but remains at a higher density than the critical density associated with the laser radiation, in order to ensure that the sample is heated entirely by X rays and not by laser light. The sample foil itself is clad in plastic (0.3 $\mu$m) to constrain the foil expansion and produce a more uniform density. The measured quantity is the transmisison of the sample, obtained from

\begin{equation}
T(h\nu)=\frac{I(h\nu)}{I_0},
\end{equation}

\noindent where $I(h\nu)$ is the attenuated intensity of the probe (backlighter) radiation and $I_0$ the reference intensity. The transmission can be related to opacity, for an homogeneous plasma in which reabsorption processes can be neglected, by Beer-Lambert-Bouguer's law:

\begin{equation}
T(h\nu)=e^{-\rho L\kappa(h\nu)},
\end{equation}

\noindent where $\rho$ is the density of the material and $L$ its thickness. Figure \ref{Al_Eidmann2+davidson3} (left) shows that the spectrum can be interpreted with a rather high accuracy (except maybe as concerns the main feature around 1530 eV) using SCO-RCG at a single temperature and a single density.

The spectrally-resolved transmission of aluminum was measured in the range of 70 to 280 eV at $\rho$=0.01 g/cm$^3$ and $T$=22 eV \cite{winhart95,winhart96}. For this purpose, the iodine laser ASTERIX IV (200 J energy with pulse duration 0.4 ns at 400 nm) was focused into a spherical gold cavity with a diameter of 3 mm. The generated radiation with a temperature $T_R\approx$ 60 eV heated thin tamped absorber foils, which were probed by the radiation of a backlighter source. In contrast to the X-ray range (around 1 keV) in which many experiments were performed using crystal spectroscopy, the spectral range below 1 keV is much less explored. It requires reliable XUV diagnostics and sufficient suppression of self-emission of the heated sample. On the other hand, it is this spectral range, which determines the Rosseland and Planck mean opacities, \emph{i.e.} the radiative transfer. As can be seen in Fig. \ref{Al_Eidmann2+davidson3} (right), the spectrum can be analyzed taking into account a temperature variation of 6 eV.

\begin{figure}[h]
\vspace{7mm}
\centering
\includegraphics[width=10cm]{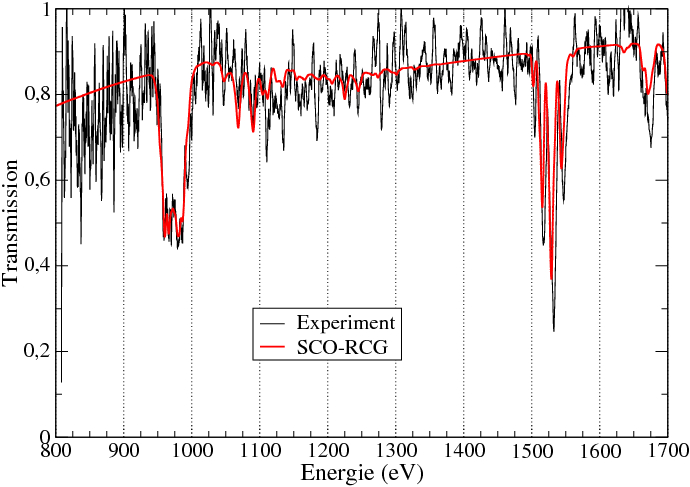}
\caption{Copper $2p-nd$, $n$=3, 4 ... (around 975 eV, 1075 eV, ...) and aluminum $1s-n'p$, $n'$=2, 3, ... (around 1530 eV, 1675 eV, ...) absorption structures measured by Dozi\`eres \emph{et al.} measured at LULI laser facility \cite{dozieres15}. Comparison between experiment and our SCO-RCG calculation at $T$=27 eV and $\rho$= 0.01 g/cm$^3$. The areal mass of copper is equal to 15 $\mu$g/cm$^{2}$ and that of aluminum to 14 $\mu$g/cm$^{2}$.}
\label{calculs_tir69}
\end{figure}

X-ray transmission spectrum of copper was measured at ``Laboratoire d'Utilisation des Lasers Intenses'' (LULI) in France at the LULI2000 laser facility with an improved design of indirect heating \cite{dozieres15}. The sample is a thin foil of mid-Z material inserted between two gold cavities heated by two 300 J, frequency-doubled (\emph{i.e.} the wavelength of the laser is $\lambda$=0.526 $\mu$m for a better X-ray conversion efficiency) nanosecond laser beams. A third laser beam irradiates a gold foil to create a spectrally continuous X-ray source (backlight) used to probe the sample. Figure \ref{calculs_tir69} shows an interpretation of the transmission of a multi-layer sample made of different materials: C (70 nm)/Al (38 nm)/Cu (12 nm)/Al (38 nm)/C (70 nm).

\begin{figure}[h]
\vspace{7mm}
\centering
\includegraphics[width=8cm]{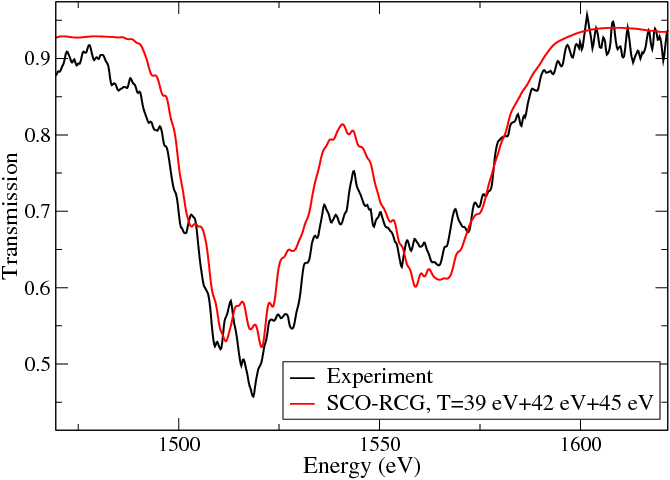}
\caption{Bromine $2p-3d$ absorption structures of an NaBr plasma measured by Bailey \emph{et al.} \cite{bailey02,bailey03}. Comparison between experiment and our SCO-RCG calculations at $n_e$=3$\times$10$^{21}$ cm$^3$, averaged over three temperatures: $T$= 39, 42 and 45 eV. The areal densities obtained from the He$_{\gamma}$ and He$_{\delta}$ Na line fits are respectively 1.0$\times$10$^{17}$ and 1.8$\times$10$^{17}$ atoms/cm$^{2}$.}
\label{BaileyT2_new_fr_2bis}
\end{figure}

There are several benefits in resorting to Z-pinch radiation for opacity measurements, including relatively large cm-scale lateral sample sizes and relatively long 3-5 ns experiments durations. Indeed, these characteristics enhance sample uniformity. The spectrally resolved transmission through a CH-tamped NaBr foil was measured by Bailey \emph{et al.} at the Sandia National Laboratories (SNL) and published in 2003 \cite{bailey03}. The Z-pinch produced the X rays for both the heating and backlight sources. There is a good agreement between observed and synthetic (SCO-RCG) spectra in the region were the $n$=2 to 3, 4 transitions in bromine ionized into the M shell are exprected to occur (see Fig. \ref{BaileyT2_new_fr_2bis}). The ratio between the two main structures ($2p_{1/2}\rightarrow 3d_{3/2}$ and $2p_{3/2}\rightarrow 3d_{5/2}$) is highly sensitive to relativistic configuration interaction (\emph{i.e.} Coulomb interaction between relativistic subconfigurations of a non-relativistic configuration). 

\begin{figure}[h]
\vspace{2mm}
\centering
\includegraphics[width=7.2cm]{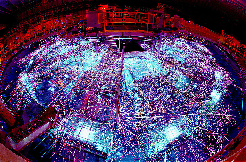}
\caption{The Z machine at SNL. \copyright Sandia National Laboratories.}
\label{Z}
\end{figure}

\begin{figure}[h]
\vspace{2mm}
\centering
\includegraphics[width=12cm]{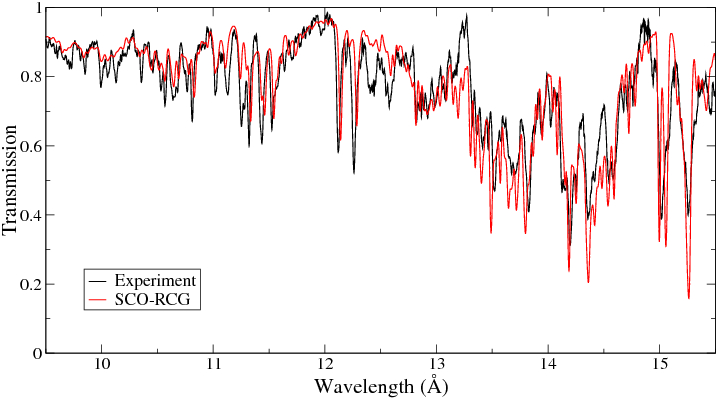}
\caption{Iron transmission spectrum measured by Bailey \emph{et al.} \cite{bailey07} on the SNL Z facility. Comparison between experiment and our SCO-RCG calculation at $T$=150 eV and $\rho$=0.058 g/cm$^3$. The areal mass, measured by Rutherford backscattering, is equal to 54 $\mu$g/cm$^{2}$.}
\label{fitBaileywithorwithoutSCO}
\end{figure}

In 2007, Bailey \emph{et al.} reported on iron transmission measurements at inferred $T$= 156 eV and $n_e$=6.9$\times$10$^{21}$ cm$^{-3}$ over the photon energy range $h\nu\approx$ 800-1800 eV \cite{bailey07}. The samples consisted of an Fe/Mg mixture fully tamped on both sides by a 10 $\mu m$ thick parylene-N (C$_8$H$_8$). The Fe/Mg mixture was fabricated by depositing 10 alternating Mg and Fe layers. The challenges of high-temperature opacity experiments were overcome here using the dynamic hohlraum X-ray source at the SNL Z facility (see Fig. \ref{Z}). The process entails accelerating an annular tungsten Z-pinch plasma radially inward onto a cylindrical low density CH$_2$ foam, launching a radiating shock propagating toward the cylinder axis. Radiation trapped by the tungsten plasma forms a hohlraum and a sample attached on the top diagnostic aperture is heated during $\approx$ 9 ns when the shock is propagating inward and the radiation temperature rises above 200 eV. The radiation at the stagnation is used to probe the sample. The experimental spectrum was well reproduced by many fine-structure opacity codes \cite{bailey07} (see Fig. \ref{fitBaileywithorwithoutSCO} for the comparison with SCO-RCG). The features around 12.4 \AA~ were not reproduced by any of the involved codes.

\vspace{2mm}

\begin{figure}[h]
\vspace{8mm}
\centering
\includegraphics[width=8cm]{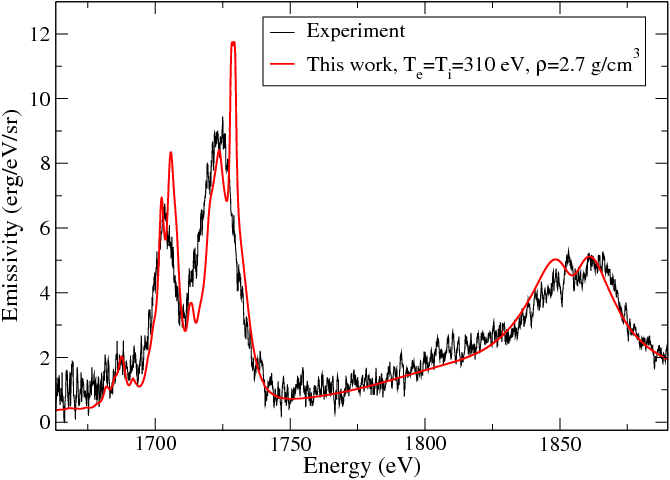}
\caption{Measured emission of aluminum ``buried layers'' heated by an ultra-short laser \cite{dervieux15} (emissive volume: 400 $\mu m^2$ $\times$ 0.5 $\mu m$, duration: 3 ps) compared to our SCO-RCG prediction at $T$=310 eV and solid density ($\rho_0$=2.7 g/cm$^3$).}
\label{fig_uhi}
\end{figure}

Hot, solid density aluminum plasmas were generated using a 4$\times$10$^{18}$ W/cm$^2$, 350 fs laser pulse at ELFIE laser facility in LULI in France \cite{dervieux15}. Ultra-high intensity (UHI) laser pulses are efficient tools to reach states of matter associated with stellar opacities. The mechanisms whereby the laser-accelerated electrons propagate and deposit their energy through the target are different from those in the ns-pulse heating. The non-thermal, high-energy electron density is usually high enough so that, in addition to the direct collisions with the target particles, the dominant heating process is the ohmic dissipation of the inductive return current formed by collisional background electrons. Time-integrated K-shell spectra and time-resolved He$_\beta$ line emission were used to infer the plasma parameters following the laser irradiation. A suite of simulation tools was employed to describe the laser-solid interaction and the subsequent radiative-hydrodynamic processes. The space-time integrated intensity spectrum is well reproduced by SCO-RCG for a mean temperature of 310 eV (see Fig. \ref{fig_uhi}). The intensity is computed here as

\begin{equation}
I(h\nu)=B(h\nu)\times\left[1-e^{-\rho L_1\kappa(h\nu)}\right]\times S_e\times\Delta t/h,
\end{equation}

\noindent where $B(h\nu)$ (erg/cm$^2$/sr) represents Planck's distribution function, $\rho$ (g/cm$^ 3$) is the density of the material, $L_1$ (cm) its reabsorption length, $S_e$ (cm$^3$) the emissive surface, $\Delta t$ (s) the emission duration and $h$ the Planck constant expressed in eV.s. The structure around 1720 eV corresponds to Ly$_{\alpha}$ and the structure around 1850 eV to He$_{\beta}$.

\vspace{6mm}

\section{Stellar envelopes}\label{env}

\subsection{$\kappa$ mechanism}

The opacity of the ``iron group'' (Cr, Fe, Ni and Cu) for $T\approx$ 2-3$\times$10$^5$ K and $\rho\approx$ 10$^{-7}$-10$^{-6}$ g/cm$^3$ is particularly important for envelopes of $\beta$ Cephei ($\beta$ Canis Majoris) stars, such as $\nu$ Eridani, $\gamma$ Pegasi, $\beta$ Crucis, $\beta$ Centauri, \emph{etc.}, which are hot blue-white stars of spectral class B. Their mass is from 8 to 20 M$_{\odot}$ and their magnitude from +3.16 to +3.27. $\beta$ Cephei are pulsating stars, progenitors of type II supernovae, and their period is about 4.57 hours. They pulsate through the $\kappa$ mechanism, due to M-shell transitions in elements of the iron group, which induce, when plotting Rosseland mean opacity as a function of $\log$(temperature in K), an opacity bump. The shape and position of this bump are very sensitive to the mass, the metallicity (fraction of mass that is not in hydrogen or helium) and the age of the considered star, showing that a precise determination of these three parameters as well as the proper knowledge of the opacities are required to understand the structure and evolution of the star. The first difficulty in interpreting their oscillation spectrum comes from the fact that some modes are observed but not predicted by stellar models.

Opacity drives the pulsations of many variable stars. In cases where the opacity increases with $T$, the atmosphere becomes unstable against pulsations. The kappa mechanism proceeds in several steps:

\vspace{2mm}

1. The inward motion of a layer of the star tends to compress the layer and increase the density $\rho$. 

\vspace{2mm}

2. The layer becomes more opaque, the flux from the deeper layers gets stuck in the high opacity ($\kappa$) region.

\vspace{2mm}

3. This heat increase causes a build-up of pressure that pushes the layer back out again. 

\vspace{2mm}

4. The layer expands, cools and becomes more transparent to radiation. 

\vspace{2mm}

5. Energy and pressure beneath the layer diminish. 

\vspace{2mm}

6. The layer falls inward and the cycle repeats. 

\vspace{2mm}

There is a renewal in the kappa mechanism studies since the launch of COROT (2006) which is the first satellite dedicated to the development of asteroseismology \cite{michel08} and search for exoplanets. This mechanism is responsible for the pulsational instability of stars between 1.5 and 20 M$_{\odot}$. In the external optically thick layers, the diffusion approximation is justified and the variation of radiation flux $F_R$ is related to luminosity $L_r$ by

\begin{equation}
\delta\mathrm{div}F_R=\frac{1}{4\pi r^2}\frac{d\delta L_r}{L_r},
\end{equation}

\noindent where

\begin{equation}
\frac{\delta L_r}{L_r}=\frac{dr}{d\ln T}-\frac{d\kappa}{\kappa}+A\left(\frac{\delta T}{T}+\frac{\delta r}{r}\right).
\end{equation}

\begin{figure}[h]
\vspace{2mm}
\centering
\includegraphics[width=6.2cm]{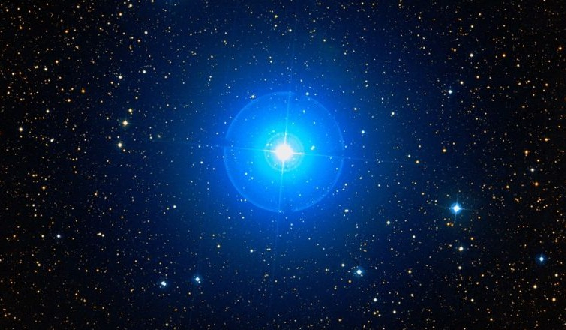}
\caption{$\beta$ Cephei star. \copyright NASA, Palomar observatory.}
\label{Alfirk}
\end{figure}

\noindent The first term describes the radiative dissipation and in fact stabilizes the star, the term $\delta T/T$ (sometimes referred to as $\gamma$ mechanism) describes the direct influence of the temperature variation on the luminosity (this effect is usually small), $\kappa$ represents the Rosseland mean opacity, $A$ is a constant, and the last term contributes to the instability because the radiating area is reduced in case of the compression \cite{turck09}. The $\kappa$ mechanism will work if the opacity varies more quickly with radius than the other terms. Two kinds of excitation are possible: the opacity bump due to the ionization of He (He II or He III), this is the so-called Eddington valve \cite{shapley14,eddington18,eddington26a,eddington26b}, at $\log T\approx$ 4.5 ($T$ in K), excites the evolved Cepheids, RR Lyrae (low mass evolved stars) and the $\delta$ Scuti stars (1.5-2.5 M$_{\odot}$ main-sequence stars) or the opacity bump is due to the M shell of iron at $\log T\approx$ 5.2, which excites the $\beta$ Cephei (see Fig. \ref{Alfirk}) stars (7-20 M$_{\odot}$ main-sequence stars), ``Slowly Pulsating B'' stars (SPB, 3-9 M$_{\odot}$ main-sequence stars) and ``B-type subdwarf'' (sdB, helium core, 0.5-1.4 M$_{\odot}$). The latter are representative of a late stage in the evolution of some stars, caused when a red giant star looses its outer layer before the core begins to fuse helium. The sdB stars are expected to become white dwarfs without going through any more giant stages. Some of them pulsate with a period from 900 to 600 seconds (such as V361 Hydrae in the Hydra constellation), and others with a period 45 to 180 minutes. Their mass is from 2 to 6 M$_{\odot}$. SPB stars are subject to gravity modes (``g'' modes), also connected to iron opacity. SPB and $\beta$ Cephei-type stars are interesting pulsators that might put important constraints on rotation and magnetic fields, two key actors for massive stars.

\subsection{Comparisons of Rosseland means}\label{ross}

The Rosseland mean opacity is the harmonic mean opacity averaged over the derivative of the Planck function with respect to the temperature:

\begin{equation}
\frac{1}{\kappa_R}=\int_0^{\infty}\frac{W_R(u)}{\kappa'(u)}du,
\end{equation}

\noindent where

\begin{equation}
u=\frac{h\nu}{k_BT}\;\;\;\; \mathrm{and}\;\;\;\; W_R(u)=\frac{15u^4e^{-u}}{4\pi^4\left[1-e^{-u}\right]^2}.
\end{equation}

\begin{figure}[h]
\vspace{5mm}
\centering
\includegraphics[width=7cm]{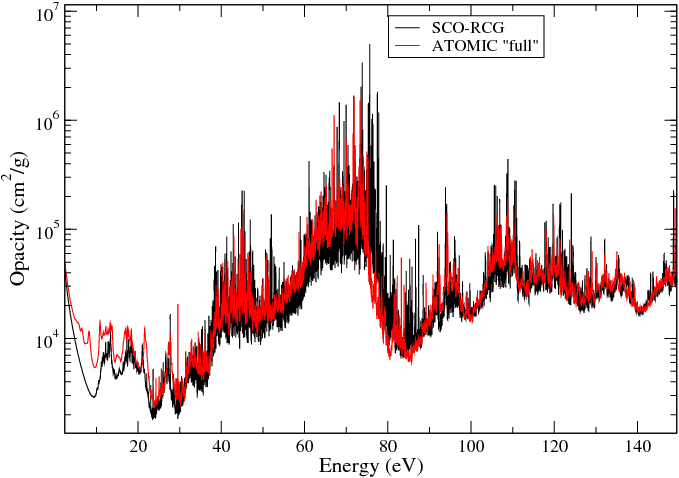}
\hspace{2mm}
\includegraphics[width=7cm]{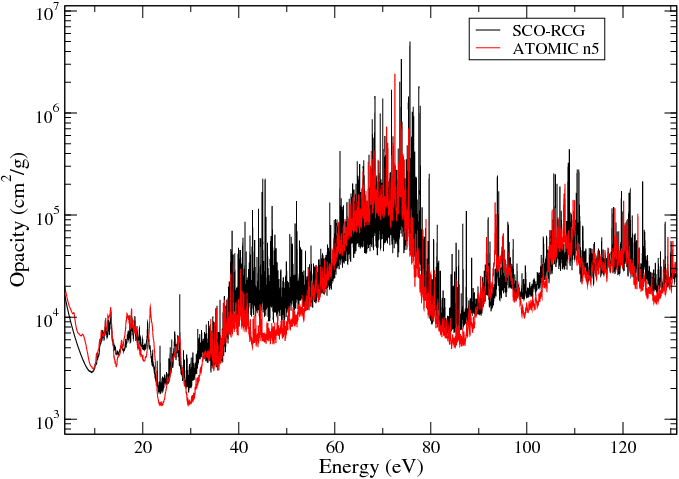}
\caption{Comparison between SCO-RCG and ATOMIC ``full'' computations \cite{fontes15,colgan16,turck16} of iron opacity at $T$=23 eV and $\rho$=2 mg/cm$^3$. The ATOMIC calculations were kindly provided by J. Colgan.}
\label{compa_2mg+compa_2mg_1}
\end{figure}

\noindent The ATOMIC-code \cite{fontes15,colgan16} ``full'' calculations allow single-configuration fine-structure detail to be included in a relatively computationally inexpensive manner. They thus include intermediate coupling within a configuration but not complete interaction among the set of configurations. Such a simplification allows one to include the contributions from many thousands of configurations. A second set of ATOMIC semi-relativistic multi-configuration calculations was also performed using a reduced list of configurations. These calculations were designed to include the transitions in the photon range of current interest ($h\nu$ lower than 250 eV), $\Delta n$=0,1, 2 and $n_{\mathrm{max}}=5$ ($n$ is the principal quantum number), referred to as ATOMIC n5. One observes that ATOMIC ``full'' and ATOMIC n5 agree within 20 \% and that, in all cases, ATOMIC n5 values are smaller than those of ATOMIC ``full''. SCO-RCG, which is not limited to $n\leq$ 5, is generally closer to ATOMIC ``full'' than to ATOMIC n5 (see Fig. \ref{compa_2mg+compa_2mg_1}, left and right, as well as tables \ref{atomic1} and \ref{atomic2}) \cite{turck16}, which is consistent with the fact that configuration interaction plays a less important role than state completeness in the present case. The values of OP (Opacity Project) \cite{op1,op2}, are usually lower. The Opacity Project, an international collaboration of about 25 scientists from 6 countries, started in response to a plea \cite{simon82}. The atomic data are available in the database TOPbase \cite{topbase} and the monochromatic opacities and Rosseland mean opacities are available at the OPServer \cite{opserver} at the Ohio Supercomputer Center.

\begin{table}[h]
 \caption{Values of Rosseland mean opacity calculated by OP, ATOMIC and SCO-RCG codes for conditions of stellar envelopes.}
\centering
\begin{tabular}{cccccc}\hline
%\toprule
\textbf{$T$ (K)}	& \textbf{$n_e$ (cm$^{-3}$)}	& \textbf{$\rho$ (g/cm$^{3}$)}	& \textbf{OP (cm$^2$/g)} & \textbf{ATOMIC ``full'' (cm$^2$/g)} & \textbf{SCO-RCG (cm$^2$/g)}\\\hline
%\midrule
125800		& 10$^{17}$			& 1.35$\times$10$^{-6}$	& 25	& 64    & 63\\
177827		& 3.16$\times$10$^{17}$	& 3.44$\times$10$^{-6}$	& 358	& 683 & 674\\
199473		& 10$^{17}$			& 9.52$\times$10$^{-7}$	& 354	& 487 & 500\\
251190		& 10$^{18}$			& 8.85$\times$10$^{-6}$	& 1270	& 1359  & 1313\\
295553		& 3.16$\times$10$^{17}$	& 2.44$\times$10$^{-6}$	& 232	& 232   & 122\\\hline
%\bottomrule
\end{tabular}
\label{atomic1}
\end{table}

\begin{table}[h]
 \caption{Rosseland mean opacities computed with two versions of ATOMIC and SCO-RCG codes at conditions that can be reached experimentally on a laser facility. These conditions are interesting since they provide an average ionization close to the one of real stellar envelopes.}
\centering
\begin{tabular}{ccccc}\hline
%\toprule
\textbf{$T$ (eV)}	& \textbf{$\rho$ (g/cm$^{3}$)}	& \textbf{ATOMIC full (cm$^2$/g)} & \textbf{ATOMIC n5 (cm$^2$/g)} & \textbf{SCO-RCG (cm$^2$/g)}\\\hline
%\midrule
21		& 2$\times$10$^{-3}$		& 19266	& 14361 & 17853\\
22		& 2$\times$10$^{-3}$		& 19613	& 14910 & 18435\\
23		& 2$\times$10$^{-3}$		& 19508	& 15205 & 18510\\
25		& 2$\times$10$^{-3}$		& 18384	& 15094 & 17550\\\hline
%\bottomrule
\end{tabular}
\label{atomic2}
\end{table}

\noindent The observed modes do not agree, as concerns their stability, with modes deduced from the OP or OPAL tables \cite{iglesias96}, which suggests that those opacities are underestimated \cite{daszynska10}. A recent study \cite{salmon12} pointed out that as much as 50 \% increase in astrophysical opacities was necessary to solve B star pulsation problems in the Magellanic cloud. Iglesias \cite{iglesias15b} showed that an increase of population I stars (metallicity of 0.02) near the iron-group bump involves an order of magnitude multiplier on the nickel opacity and more than an overall doubling of the iron opacity. Although there are uncertainties in theoretical opacities (spectral-line models, configuration-interaction accounting, mixture models, ...), these errors seem unable to explain such large opacity increases. Furthermore, code comparisons do show differences in spectral details, but also suggest that current advanced implementations of photon-absorption theories are unlikely to produce large overall discrepancies in the iron-group opacities for matter conditions relevant to B stars. 

Moravveji \cite{moravveji15} produced new opacity tables with enhanced iron and nickel contributions to the Rosseland mean opacity by 75 \% each, and computed, with the MESA (Modules for Experiments in Stellar Astrophysics) code \cite{paxton15}, three dense grids of evolutionary models for Galactic O- and B-type stars covering from 2.5 to 25 M$_{\odot}$ from zero-age main sequence until effective temperature $T_{\mathrm{eff}}$=10000 K after the core hydrogen exhaustion. The authors carried out non-adiabatic mode stability analysis and found that $\approx$ 75 \% enhancement, only in the iron opacity, is sufficient to consistently reproduce the observed position of late O-type $\beta$ Cepheids and eight hybrid B-type pulsators on the Kiel diagram, \emph{i.e.} $\log T_{\mathrm{eff}}$ vs $\log g$, where $g$ represents the stellar surface gravity \cite{langer14}.

\section{The boundary of the radiative / convective zones of the Sun}\label{bou}

\begin{figure}[h]
\vspace{4mm}
\centering
\includegraphics[width=7cm]{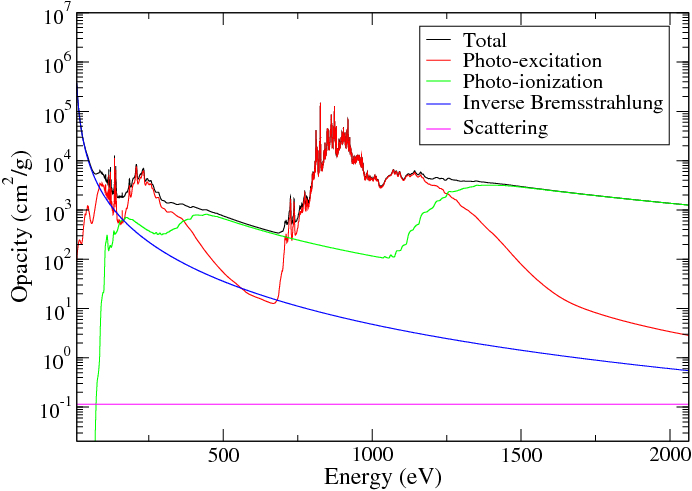}
\hspace{2mm}
\includegraphics[width=7cm]{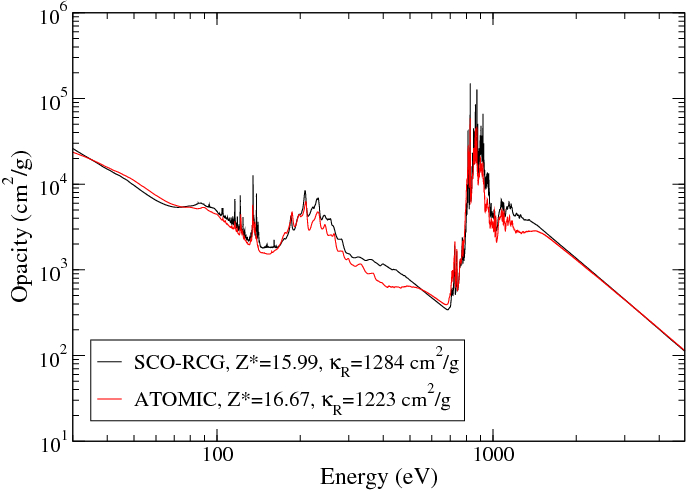}
\caption{Different contributions to the opacity of iron calculated with SCO-RCG at conditions of the boundary of the convective zone (left), {\em \emph{i.e.}} $T$=192.91 eV and $n_e$=10$^{23}$ cm$^{-3}$, and a comparison between the present SCO-RCG and ATOMIC total opacities \cite{fontes15,colgan16} at the same conditions (right). The ATOMIC calculations were kindly provided by J. Colgan. The OP Rosseland mean at these conditions is much lower and equal to 854 cm$^2$/g.}
\label{contrib_bcz+sco-rcg_bcz}
\end{figure}

The Sun is our closest star and is thus used as a benchmark to study other stars. Its radius, luminosity and mass are known with a great accuracy. However, some doubts are raised on the relevance of the used microscopic physics. In the past decade, the photospheric abundances of the Sun have been revised several times by many observers \cite{asplund05,asplund09,yang16}. The standard solar models constructed with the new low-$Z$-metal abundances disagree with helioseismic results and detected neutrino fluxes \cite{turck01a,turck01b,antia06,guzik08,serenelli09,turck11}. For instance, a discrepancy between helioseismic observations and predictions by standard solar models appeared in the solar sound speed profile (sound speed versus radius). This discrepancy, of about 20 times the vertical error bar, raises some questions on the solar radiative transfer. In addition, the Sun has an inner radiative heat conduction zone that gives way to a convective zone nearer the surface. The different contributions to the opacity of iron calculated with SCO-RCG at conditions of the boundary of the convective zone (BCZ) are displayed in Fig. \ref{contrib_bcz+sco-rcg_bcz} (left) as well as a comparison between our SCO-RCG and ATOMIC codes (right). The Rosseland means are rather close, although some differences exist around $h\nu$=800 eV (maximum of the derivative of Planck's distribution with respect to temperature). 

Solar models typically find a location of the boundary between these zones that differs significantly from the measured one (by more than 13 standard deviations \cite{basu08}). Several hypotheses can explain this difference:

\vspace{2mm}

$\bullet$ macroscopic processes in the radiative zone are not taken into account in the energetic balance of the Sun.

\vspace{2mm}

$\bullet$ the radiative transfer calculations are not accurate, either as concerns the Rosseland mean value that could be underestimated or in the treatment of the radiative acceleration which limits the gravitational settling and could lead to incorrect internal abundances. The radiative acceleration of species $k$ reads

\begin{equation}
g(k)=\frac{F_R}{c}\frac{M}{M_k}\kappa_R\gamma(k),
\end{equation}

\noindent where $F_R$ is the radiation flux, $M$ the total mass of the star, $M_k$ the mass of species $k$ and

\begin{equation}
\gamma(k)=\int\frac{\kappa_{\nu}(k)}{\kappa_{\nu}(\mathrm{total})}f_{\nu}d\nu,
\end{equation}

\noindent where $\kappa_{\nu}(k)$ is the monochromatic opacity of species $k$ and $f_{\nu}$ is a function of the frequency and of the temperature. The unexplained discrepancies could also be due to all these effects simultaneously. The heavy elements significantly contribute to opacity even if they are present only at few percents in mass fraction in the solar mixture, which is mainly constituted of hydrogen and helium. The most important contributors are:

\vspace{2mm}

$\bullet$ iron, which contributes to the total opacity (including hydrogen and helium) at a level of 20 \% in most of the radiative zone because it is always partially ionized;

\vspace{2mm}

$\bullet$ oxygen, which becomes partially ionized at 0.6 R$_{\odot}$ and plays a major role at the basis of the convective zone. The increase of its opacity contribution triggers the convection;

\vspace{2mm}

$\bullet$ silicon, which contributes about 10 \% at temperatures below 10 MK.

\vspace{2mm}

Rotation, enhanced diffusion, convection overshoot and magnetic field contribute to explain the discrepancy. The too low helium subsurface abundance in enhanced diffusion models can be improved by the mixing caused by rotation and magnetic fields \cite{yang16}. The problem of the depth of the convective zone in rotating models can be resolved by accounting for convection overshoot. The latest phenomenon refers to convection carrying matter beyond and unstable region into a stratified, stable region. Overshoot is due to the momentum of the convective material, which carries the material beyond the unstable region. The heat of the Sun's thermonuclear fusion is carried outward by radiation in the deep interior radiation zone and by convective circulation in the outer convection zone. However, cool sinking matter from the surface penetrates farther into the radiative zone than theory would suggest. The Asplund-Grevesse-Sauval rotation model \cite{asplund09} including overshooting reproduces the seismically inferred sound-speed and density profiles and the convection zone depth, but fails to reproduce the surface helium abundance and neutrino fluxes.

More generally, the solar abundance controversy inspires many investigations in the microscopic physics. For instance, Mussack and D\"appen examined the correction to the proton-proton reaction rate due to dynamic screening effects \cite{mussack11}.

Stars generate low-mass weakly interacting particles which are responsible for energy losses, such as the axions, emitted by the Primakoff effect \cite{schlattl99}. Axions are elementary particles postulated in 1977 to solve the strong CP problem in quantum chromodynamics. They are a possible component of cold dark matter. It was reported in 2014 that axions have been detected as a seasonal variation in observed X-ray emission that would be expected from conversion in the Earth's magnetic field of axions streaming from the Sun \cite{fraser14}. In non-hadronic axion models, where axions couple to electrons, the solar axion flux is completely dominated by the ABC reactions (Atomic recombination and deexcitation, Bremsstrahlung and Compton). The ABC flux was computed from available libraries of monochromatic photon radiative opacities by exploiting the relations between axion and photon emission cross sections. These results turn to be $\approx$ 30 \% larger than previous estimates due to atomic recombination (free-bound electron transitions) and deexcitation (bound-bound), which where not taken into account \cite{redondo13}.

The spectra of silicon, in particular with high resolution, are helpful for deducing the abundance of silicon in stars. In the past decades, the K-shell absorption lines of silicon ions in various astrophysical objects have been extensively observed with high-resolution spectrometers of the XMM-Newton, Chandra, and Suzaku space missions. A typical example of a spectrometer is the High-energy Transmission Grating Spectrometer (HETGS) on board the Chandra X-ray Observatory. With the HETGS, silicon absorption lines were observed in the spiral galaxy NGC 3783, in the black hole candidate Cygnus X-1, in the ultracompact X-ray Dipper 4U 1916-05, in the Seyfert 1 galaxy NGC 5548, in the bright Seyfert 1 galaxy MCG 6-30-15 and NGC 3516, in the Seyfert 1 active galactic nucleus NGC4051, \emph{etc.} \cite{xiong16}. Silicates are an important component of cosmic matter \cite{savin12}. They form in the winds of AGB (asymptotic giant branch) stars. The AGB is a region of the Hertzsprung-Russell diagram populated by evolved cool luminous stars. This is a period of stellar evolution that is experienced by all low- to intermediate-mass stars (0.6-10 M$_{\odot}$) late in their lives. An AGB star appears, for a large part of its life, as a bright red giant with a luminosity thousands of times greater than the Sun. Silicates are processed in the diffuse interstellar medium. They are also present in dust of protoplanetary and debris disks where they help regulating thermal exchanges. In circumstellar environments, evidence for crystalline silicates is found both around AGB stars and in disks around Ae/Be stars, T Tauri stars and brown dwarfs \cite{savin12}. They are also found in cometary environments, in asteroid spectra and in interplanetary dust particles. Accurate interpretation of their absorption lines requires detailed theoretical models \cite{wei08}. 

\begin{table}[h]
 \caption{Temperature and free-electron density at different depths of the Standard Solar Model computed with the MESA code \cite{paxton15} for the Asplund \emph{et al.} composition \cite{asplund09}.}
\centering
\begin{tabular}{ccc}\hline
%\toprule
\textbf{Solar radius fraction ($r/R_{\odot}$)}	& \textbf{$T$ (eV)}	& \textbf{$n_e$ (cm$^ {-3}$)}\\\hline
%\midrule
0.5		& 340			& 8$\times$10$^{23}$\\
0.6		& 270			& 2.5$\times$10$^{23}$\\
0.7		& 200			& 10$^{23}$\\\hline
%\bottomrule
\end{tabular}
\label{radius}
\end{table}

\begin{figure}[h]
\vspace{6mm}
\centering
\includegraphics[width=8cm]{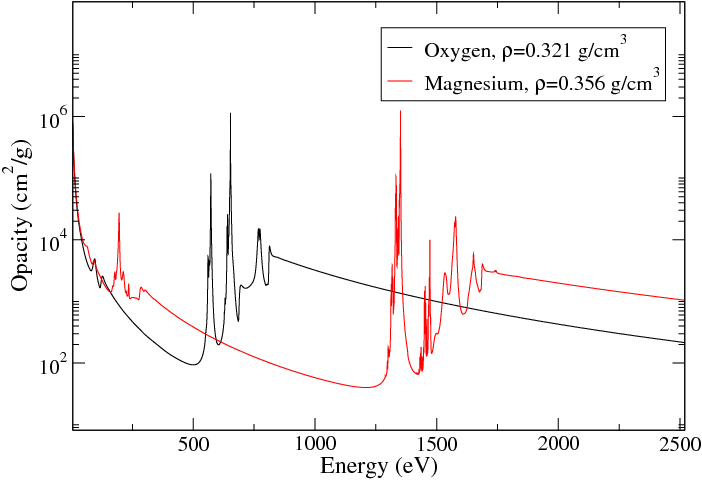}
\caption{Opacity of oxygen and magnesium computed by SCO-RCG for $T$=174.92 eV and a total density of the mixture of $\rho$=0.16 g/cm$^3$ (the partial densities of oxygen and magnesium are 0.321 and 0.356 g/cm$^3$ respectively), corresponding to solar radius $R$=0.726.}
\label{oxygen-magnesium}
\end{figure}

Oxygen and magnesium are also important for the modelling of the Sun (their respective opacities are shown in Fig. \ref{oxygen-magnesium}).

Reproducing the solar interior is a real challenge because one tries to reproduce the charge state distribution of the different elements together with the free-electron density $n_e$ at the targeted conditions, that means density greater than that of a solid and high temperature (see table \ref{radius}). 

\section{Attempts to understand the enigmatic photo-absorption experiment on iron performed by Bailey \emph{et al.} at Sandia National Laboratory}\label{bai}

As mentioned in Section \ref{int}, Bailey \emph{et al.} published in 2007 a measurement, performed on the Z-pinch facility, of the iron transmission at conditions lower than those of the basis of the convective zone ($T$=156 eV and $\rho$=0.058 g/cm$^3$) \cite{bailey07}. This first measurement agrees well with most of the theoretical predictions. Then they have increased the temperature up to 196 eV and reached free-electron densities of several times 10$^{21}$ cm$^{-3}$. However, for this last measurement, an unexplained gap exists between the measurement and the theoretical calculations \cite{bailey15}. 

\subsection{Effect of density and temperature}

\begin{figure}[h]
\vspace{6mm}
\centering
\includegraphics[width=8cm]{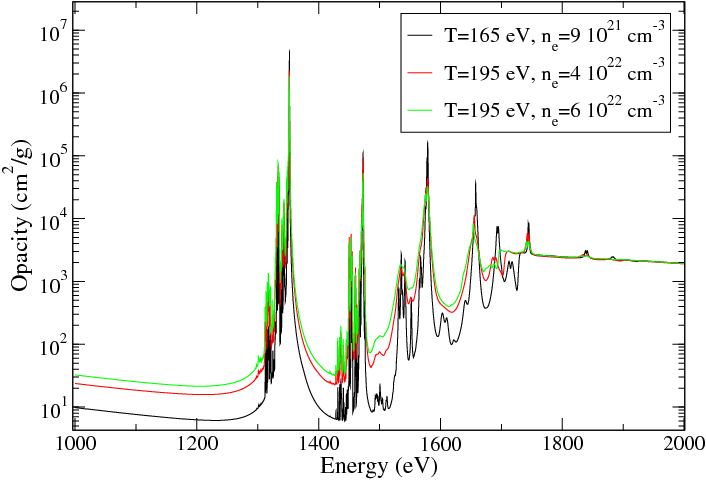}
\caption{Magnesium opacity computed by SCO-RCG at $T$=165 eV, $n_e$=9$\times$10$^{21}$ cm$^{-3}$, $T$=195 eV, $n_e$=4$\times$10$^{22}$ cm$^{-3}$ and $T$=195 eV, $n_e$=6$\times$10$^{22}$ cm$^{-3}$.}
\label{fig_Mg_tn}
\end{figure}

Density and temperature were inferred by Nagayama \emph{et al.} \cite{nagayama16}, in the experiment described in Ref. \cite{bailey15}, by analysis of Stark profiles of magnesium K-shell lines. As shown in Fig. \ref{fig_Mg_tn}, the line profiles (intensity, asymmetry, wings, ...) are very sensitive to the plasma conditions, and can therefore be used to diagnose the plasma temperature and density. In the SNL experiment, the analysis of the K-shell transmission spectra of Mg leads to $T$=182 eV and $n_e$=10$^{22}$ cm$^{-3}$ (which corresponds roughly to $\rho$=0.17 g/cm$^3$) \cite{nagayama16}.

\begin{figure}[h]
\vspace{2mm}
\centering
\includegraphics[width=7.5cm]{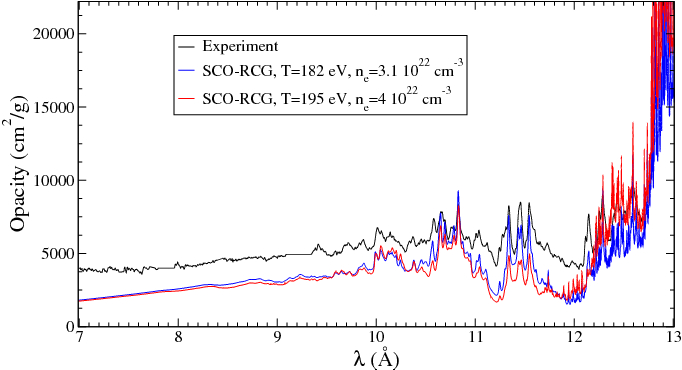}
\hspace{2mm}
\includegraphics[width=7.5cm]{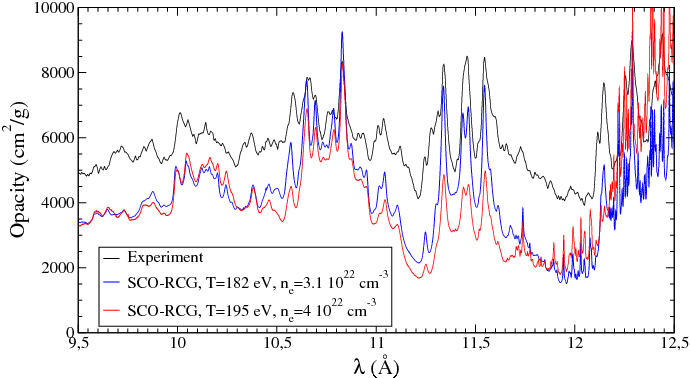}
\caption{Opacity computed by SCO-RCG in two different conditions: $T$=182 eV, $n_e$=3.1$\times$10$^{22}$ cm$^{-3}$ and $T$=195 eV, $n_e$=4$\times$10$^{22}$ cm$^{-3}$.}
\label{fig1+2}
\end{figure}

Figure \ref{fig1+2} displays the opacity computed by SCO-RCG in two different conditions: $T$=182 eV, $n_e$=3.1$\times$10$^{22}$ cm$^{-3}$ and $T$=195 eV, $n_e$=4$\times$10$^{22}$ cm$^{-3}$. The amplitude of the structures is better reproduced at $T$=182 eV, $n_e$=3.1$\times$10$^{22}$ cm$^{-3}$, but the predicted opacity level is much lower than the experimental one in both cases. However, the highest opacity values (peaks around 10.7 and 11.5 \AA) at $T$=182 eV, $n_e$=3.1$\times$10$^{22}$ cm$^{-3}$ are consistent with the experimental ones.

\begin{figure}[h]
\vspace{2mm}
\centering
\includegraphics[width=8cm]{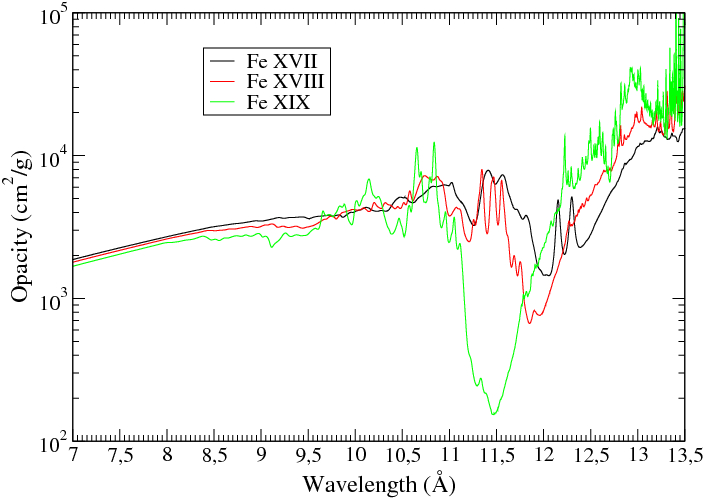}
\caption{Opacity of the dominant ions in the spectrum Fe XVII, Fe XVIII and Fe XIX at $T$=182 eV and $n_e$=3.1$\times$10$^{22}$ cm$^{-3}$ ($\approx$ 0.17 g/cm$^3$) computed by SCO-RCG. The opacity of each charge state is not weighted by the ionic fraction.}
\label{ions}
\end{figure}

The dominant ions in the spectrum (see Fig. \ref{ions}) at $T$=182 eV and $n_e$=3.1$\times$10$^{22}$ cm$^{-3}$ are Fe XVII (24.9 \%), Fe XVIII (33 \%) and Fe XIX (21.7 \%).

\begin{figure}[h]
\vspace{3mm}
\centering
\includegraphics[width=8cm]{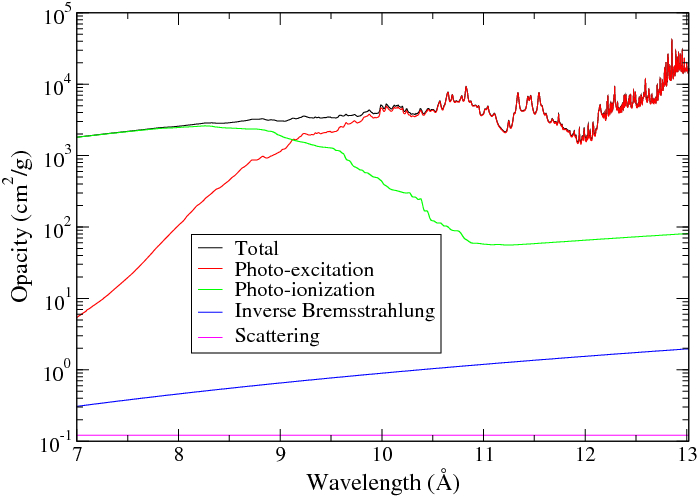}
\caption{Contributions to the total opacity at $T$=182 eV and $n_e$=3.1$\times$10$^{22}$ cm$^{-3}$ ($\approx$ 0.17 g/cm$^3$) computed by SCO-RCG.}
\label{contributions}
\end{figure}

The contributions to the total opacity (photo-excitation, photo-ionization, inverse Bremsstrahlung and scattering of a photon by a free electron) at $T$=182 eV and $n_e$=3.1$\times$10$^{22}$ cm$^{-3}$ are represented in Fig. \ref{contributions}. The most important processes are photo-excitation and photo-ionization.

%\vspace{2mm}

\begin{figure}[h]
\centering
\includegraphics[width=11.5cm]{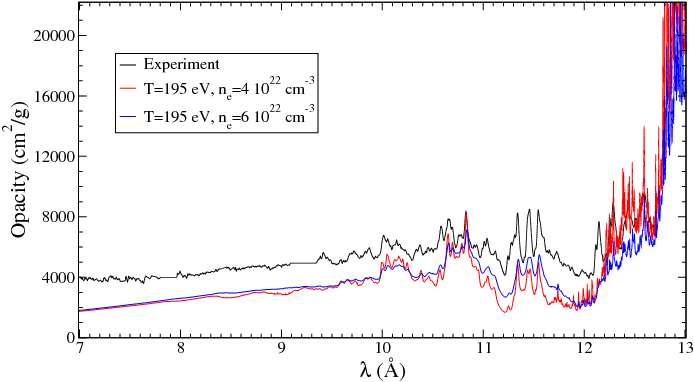}
\hspace{10mm}
\includegraphics[width=11.5cm]{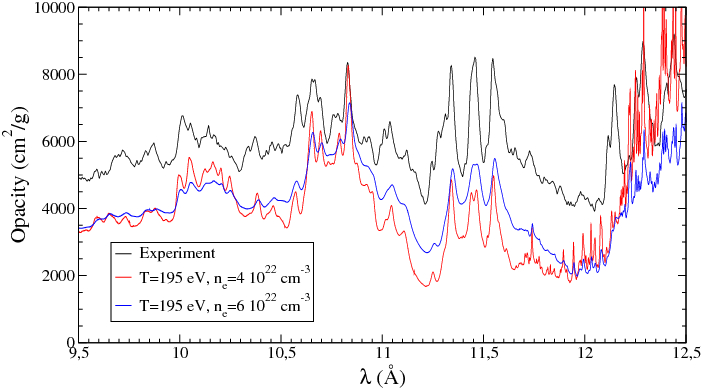}
\caption{Effect of density on the iron opacity at $T$=195 eV (SCO-RCG computations). The experimental spectrum is also represented \cite{bailey15}.}
\label{Fe_195eV+zoom}
\end{figure}

Figure \ref{Fe_195eV+zoom} shows the effect of density on the computed spectrum at $T$=195 eV. At the highest density ($n_e$=6$\times$10$^{22}$ cm$^{-3}$), the gaps between the structures (for instance between the three F-like 2p-4d features around 11.5 \AA), are less pronounced than at the lower electron density $n_e$=4$\times$10$^{22}$ cm$^{-3}$ and the wings of lines are broader, exhibiting shoulders. However, even at the highest density, the discrepancy between theory and experiment remains prominent. 

\vspace{2mm}
 
\subsection{Effect of highly excited states}

In multiply-charged ion plasmas, a significant number of electrons may occupy high-energy orbitals. These ``Rydberg'' electrons, when they act as spectators, are responsible for a number of satellites of X-ray absorption or emission lines, yielding an effective broadening of the resonance lines (red wings). The contribution of such satellite lines may be important, because of the high degeneracy of the relevant excited configurations which give a large Boltzmann weight. However, it is in general difficult to take those configurations into account since they are likely to give rise to a large number of lines. We proposed to model the perturbation induced by the spectators in a detailed way, inspired by the Partially Resolved Transition Array approach recently published by Iglesias \cite{iglesias12a,iglesias12c}, and extended to the STA theory \cite{pain15a,wilson15}. This approach, based on the additivity of the variances from active and spectator electrons in the STA theory, consists in a reduced detailed-line-accounting calculation, omitting these Rydberg electrons, the effect of the latests being included through an additional shift and broadening of the lines, expressed in terms of canonical partition functions, key-ingredients of the STA model. The resulting method can \emph{a priori} be used in any detailed-configuration/line-accounting opacity code. 

\begin{figure}[h]
\centering
\includegraphics[width=7.5cm]{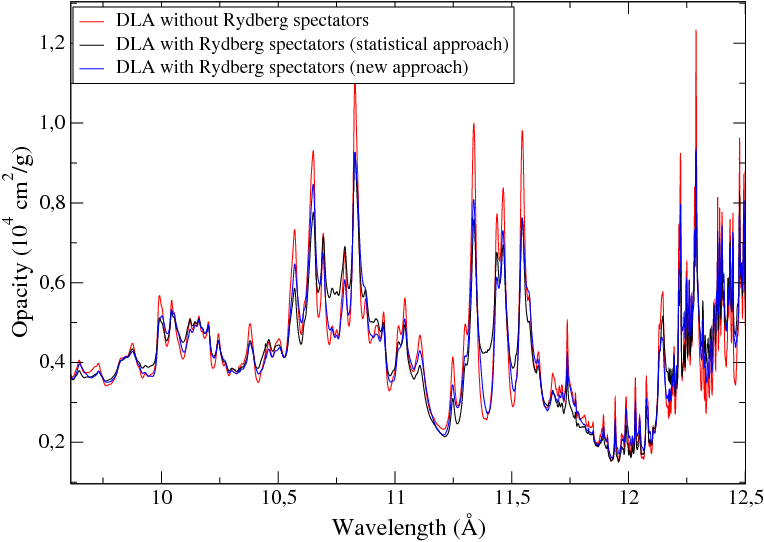}
\hspace{2mm}
\includegraphics[width=7.5cm]{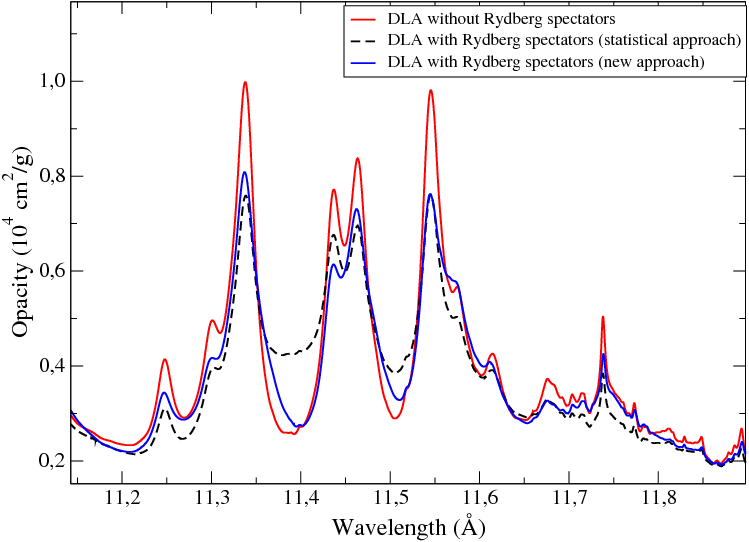}
\caption{Modelling of satellites [Be] 2p$^5$(5s $\cdots$ 8d)$^1\rightarrow$ 2p$^4$4d$^1$(5s $\cdots$ 8d)$^1$ with SCO-RCG code.}
\label{figure_loisel2014_quad}
\end{figure}

In the case of an iron plasma at $T$=182 eV and $n_e$=10$^{22}$ cm$^{-3}$ (conditions of the recent experiment by Bailey \emph{et al.} \cite{bailey15}), the statistical modelling of satellites [Be] $2p^5(5s\cdots 8d)^1\rightarrow 2p^44d^1(5s\cdots 8d)^1$ fills significantly the gaps around 11.4 and 11.5 \AA~ (see Fig. \ref{figure_loisel2014_quad}). The total contribution of satellites (blue curve) is very similar to the resonance transition (red curve), except that it is slightly shifted and broadened. These spectator electrons are very weakly bound to the ion, so that they do not interact much with the core electrons and perturb very weakly the transition. In our previous approach, the satellite lines due to the Rydberg electrons were treated in the STA theory, and yielded an overestimation of the opacity in the gap between the three main features of the F-like $2p\rightarrow 4d$ lines.

\clearpage

\vspace{2mm}

\subsection{Comparison to cold opacity}

\begin{figure}[h]
\vspace{6mm}
\centering
\includegraphics[width=8cm]{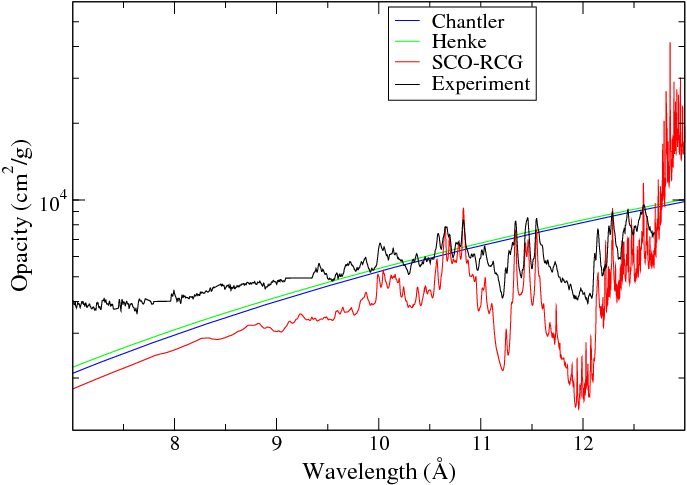}
\caption{Comparison between cold opacities published by Henke \emph{et al.} \cite{henke93} and Chantler \cite{chantler95} and the experimental spectrum \cite{bailey15}. The SCO-RCG opacity at $T$=182 eV and $n_e$=3.1$\times$10$^{22}$ cm$^{-3}$ is also represented.}
\label{cold_opacity}
\end{figure}

Figure \ref{cold_opacity} displays a comparison betweeen the experimental spectrum \cite{bailey15} and two cold opacities published by Henke \emph{et al.} \cite{henke93} and Chantler \cite{chantler95}.

As was also pointed out by Iglesias \cite{iglesias15a}, recalling that the oscillator strength density is conserved, a smoothed cross-section obtained by the following convolution is constrained to be smaller than the cold opacity:

\begin{equation}
\kappa(\nu)=\int_{-\infty}^{\infty}d\nu'G\left(\nu-\nu'\right)\kappa\left(\nu'\right)\leq\kappa_{\mathrm{cold}},
\end{equation}

\noindent where $G$ represents physical (Doppler, Stark, Van der Waals neutral-atom collisions, autoionization) and instrumental broadenings. The largest theoretical opacity is obtained from ions with a full L shell, which reproduces the cold data at small wavelength. The analysis revealed other puzzling aspects of the data: the integrated absorption suggests a full L shell, inconsistent with the measured spectrum, which displays strong spectral lines from F- and O-like configurations, and the average slope of the absorption disagrees with the one of the cold data $\kappa_{\mathrm{cold}}$. Clearly, the Sandia data exceed $\kappa_{\mathrm{cold}}$ at lower wavelengths. 

\subsection{R-matrix photo-ionization}

\begin{figure}[h]
\vspace{6mm}
\centering
\includegraphics[width=8cm]{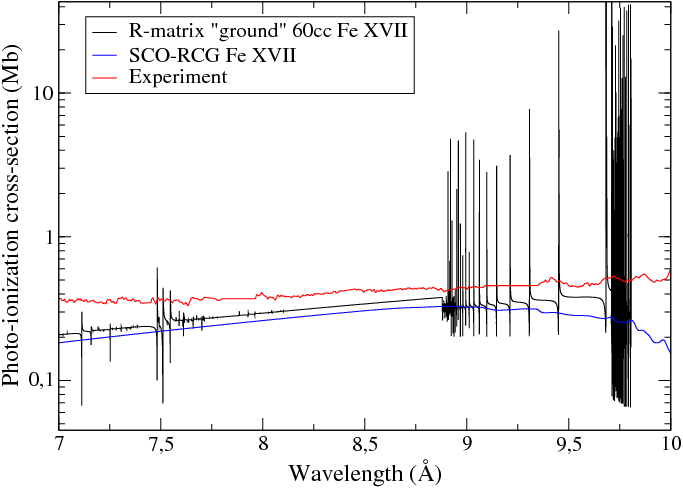}
\caption{Comparison between the present SCO-RCG (distorted-wave) in blue and R-matrix \cite{nahar11} photo-ionization cross-sections (black curve) obtained from \cite{topbase}. The experimental spectrum is also represented (red curve). The resonances in the black curve (PEC: photo-excitation of core) are the signature of transitions to a state coupled to the continuum (also known as Beutler-Fano autoionizing resonances \cite{cowan81}).}
\label{comp_1}
\end{figure}

Recently \cite{nahar16}, Nahar \emph{et al.} reported extensive R-matrix calculations \cite{nahar11} of unprecedented complexity for iron ion Fe XVII, with a wavefunction expansion of 99 Fe XVIII LS core states from $n\leq 4$ complexes (equivalent to 218 fine-structure levels) and found a large enhancement in background photo-ionization cross-sections (up to orders of magnitude) in addition to strongly peaked photo-excitation-of-core resonances (see the black curve of Fig. \ref{comp_1}).

\begin{table}[h]
 \caption{$\kappa_R$ calculated with different opacity codes relative to experiment.}
\centering
\begin{tabular}{cc}\hline
%\toprule
\textbf{Source}	& \textbf{$\kappa_R$ relative to experiment} \\\hline
%\midrule
OP \cite{op1,op2}			& 0.59 \\
R-matrix \cite{nahar11,nahar16}   & 0.51 \\
ATOMIC 	\cite{fontes15,colgan16}	& 0.60 \\
OPAS \cite{blancard12,mondet15,lepennec15}		& 0.70 \\
SCO-RCG	\cite{pain15a}	& 0.64 \\
SCRAM \cite{hansen07}		& 0.77 \\
TOPAZ \cite{iglesias03,iglesias04}		& 0.62 \\
Cold \cite{henke93}		& 0.75 \\\hline
%\bottomrule
\end{tabular}
\label{rmatrix}
\end{table}

\begin{figure}[h]
\vspace{8mm}
\centering
\includegraphics[width=8cm]{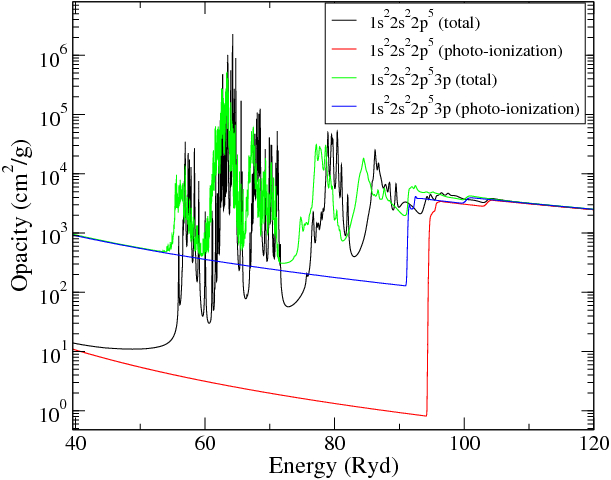}
\caption{Total and photo-ionization opacities of configurations $1s^22s^22p^5$ and $1s^22s^22p^53p$ (SCO-RCG calculations).}
\label{iglesias_wings}
\end{figure}

Figure \ref{iglesias_wings} displays total and photo-ionization opacities of ground $1s^22s^22p^5$ and excited $1s^22s^22p^53p$ configurations. As pointed out by Iglesias \cite{iglesias16}, it clearly illustrates the fact that the photo-ionization from $3p$ is important and must absolutely be taken into account in the opacity. The point here is that the latest contribution is included in all opacity codes relying on a distorted-wave modelling of photo-ionization, but was not included in the previous R-matrix computations.

It appears \cite{iglesias16}, that major opacity models already include cross-sections that are equivalent to the enhancements reported by the R-matrix method. The fact that the R-matrix calculations neglected important cross-sections (\emph{e.g.} photo-ionization involving $3p$ subshell, as mentioned above) explains why the resultant opacity is lower than other models in the spectral range measured in transmission experiments relevant to the solar interior (see table \ref{rmatrix}). The missing processes help understand the lower Fe XVII R-matrix opacity relative to other models making it discrepant with the SNL measurements \cite{blancard16}. The opacities from all models are larger than R-matrix over the whole energy range of the experiment except for a few narrow features. Moreover, they all reasonably agree, except R-matrix, at higher energies corresponding to the L-shell photo-ionization continuum. The R-matrix calculations consider photo-absorption for 283 LS terms of Fe XVII corresponding to the configurations

\begin{equation}
\begin{array}{l}
2s^22p^5n\ell\;\;\;\;\mathrm{for}\;\;\;\; n=2-10\\
2s^22p^6n\ell\;\;\;\;\mathrm{for}\;\;\;\; n=3,4\;\;\;\;\mathrm{plus}\;\;\;\; n\ell=5s
\end{array}
\end{equation}

\noindent excluding levels with $J>8$. For the target ion, R-matrix calculations includes the Fe XVIII configurations

\begin{equation}
\begin{array}{l}
2s^22p^5\\
2s^22p^4n\ell\;\;\;\;\mathrm{for}\;\;\;\; n=3,4\\
2s2p^6\\
2s2p^5n\ell\;\;\;\;\mathrm{for}\;\;\;\; n=3 \;\;\;\;\mathrm{plus}\;\;\;\; n\ell=4s,4p\\
\end{array}
\end{equation}

\noindent which generate 99 LS-terms. It follows that Nahar and Pradhan omitted the photo-ionizations

\begin{equation}
\begin{array}{ll}
2s^22p^5n\ell+h\nu\rightarrow 2s^22p^5n\ell\epsilon p, & n\ell=4p,4f \;\;\;\;\mathrm{and}\;\;\;\; n\ge 5\\
2s2p^6n\ell+h\nu\rightarrow 2p^6n\ell\epsilon p, & \mathrm{all}\;\;\;\; n\\
2s^22p^5n\ell+h\nu\rightarrow 2s2p^5n\ell\epsilon\ell', & n\ge 5\\
2s2p^6n\ell+h\nu\rightarrow 2s2p^5n\ell\epsilon\ell', & n\ell=4p,4f \;\;\;\;\mathrm{and}\;\;\;\; n\ge 5.
\end{array}
\end{equation}

\noindent with $\ell'=s,d$. Opacity models like ours (relying on distorted-wave approximation) treat the absorption processes described by Nahar and Pradhan as inner-shell bound-bound transitions into discrete autoionizing levels or as bound-free transitions. It is emphasized that all approaches satisfying the $f$-sum rule \cite{thomas25,kuhn25,cowan81} yield the same total absorption strength but only if they use the same set of initial and final configurations. The Thomas-Reiche-Kuhn sum rule reads:

\begin{equation}
\sum_{i,j}f_{ij}+\sum_i\int_{I_i}^ {\infty}\frac{df_{i,\epsilon}}{d\epsilon}d\epsilon=Z-Z^*,
\end{equation}

\noindent where $I_i$ is the ionization potential of subshell $i$ and $Z^*$ the average ionization ($Z-Z^*$ representing therefore the total number of bound electrons). Differences in the opacity can still occur due to several aspects of atomic-structure calculations, such as level populations, and/or details in broadening of spectral lines or photo-ionization resonances. Nahar and Pradhan's R-matrix calculations, however, did not include all target ion states consistent with their initial configurations and thus neglected important cross-sections. The R-matrix calculations also limited the initial configuration list compared with current opacity models. Some of the missing transitions impact the L-shell photo-ionization continuum and help explain the lower opacity from R-matrix relative to other models in the spectral range of the SNL experiments. Effects from other missing atomic data involve numerous weak features in spectral regions of low photo-absorption, such as PEC (photo-excitation-of-core) resonances. Proponents of close-coupling methods argue that such calculations are in principle better than other schemes (e.g.; distorted wave), but that may not be true in the plasma environment. In plasmas, fluctuating electric fields and collisions between ions and electrons may significantly impact, and even destroy, the coherence required for the interference effects included in close-coupling methods. It remains an open question if the improvements in cross-sections reported using R-matrix methods for isolated atoms are really relevant for the modelling of atomic processes in plasmas \cite{iglesias16,nahar16b}.

\subsection{Autoionization}

The process of autoionization, described in Sec. \ref{rad}, is likely to provide an additional broadening to spectral lines. The autoionization rate 

\begin{equation}
A_{if}^a=\sum_j\left|\langle\psi_f,j;J_T,M_T\left|\sum_{i<j}\frac{1}{r_{ij}}\right|\psi_i\rangle\right|^2,
\end{equation}

\noindent where $J_T$ and $M_T$ are the total angular momentum and its projection, $j$ the quantum number of the free electron, $\psi_i$ and $\psi_f$ being the wavefunctions of the initial and final configurations respectively. The autoionization rate can be calculated as a configuration average \cite{sampson09,gao10}. We developed such an approach in the Multi-Configuration Dirac-Fock (MCDF) code developed by Bruneau \cite{bruneau83,bruneau84}, and our results are very close to the ones obtained with the Flexible Atomic Code (FAC) \cite{gu08}.

\begin{figure}[h]
\vspace{4mm}
\centering
\includegraphics[width=8cm]{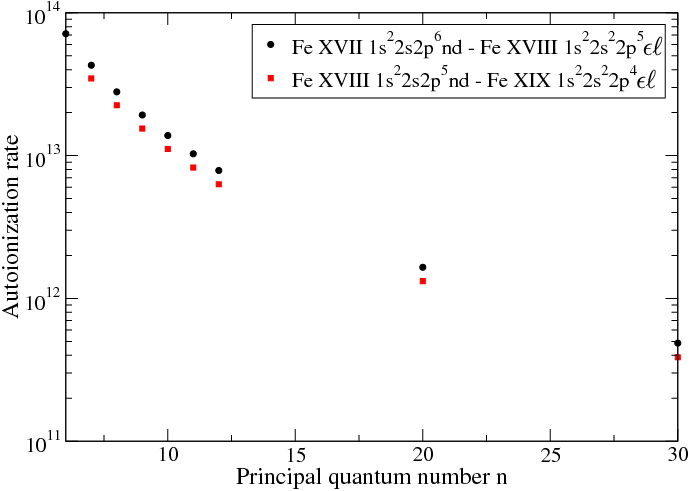}
\caption{Autoionization rate for Fe XVII $1s^22s^12p^6nd$ $\rightarrow$ Fe XVIII $1s^22s^22p^5\epsilon\ell$ and Fe XVIII $1s^22s^12p^5nd$ $\rightarrow$ Fe XIX $1s^22s^22p^4\epsilon\ell$, $n$=1, 20.}
\label{auger_hansen}
\end{figure}

High populations in metastable states can also contribute to a process called ladder ionization, which becomes important when the electron temperature is too small to support significant collisional ionization from the ground state but large enough to support ionization from excited states that lie much closer to the continuum limit. The collisional ionization flux and charge state distribution therefore depend on the degree of state detail. In addition, some mid-shell ions support excitation-ionization processes for states formed by single excitation from inner subshells, such as the $1s^ 22s^22p^ 6n\ell$ states in F-like iron, which are autoionizing for $n\ge 6$ \cite{ralchenko16}. The values of the autoionization rates for Fe XVII $1s^ 22s^12p^6nd$ $\rightarrow$ Fe XVIII $1s^22s^22p^5\epsilon\ell$ and Fe XVIII $1s^22s^12p^5nd$ $\rightarrow$ Fe XIX $1s^22s^22p^4\epsilon\ell$, for $n$=1 to 20, are plotted in figure \ref{auger_hansen}. The rates can reach several 10$^ {13}$ s$^ {-1}$, which is at maximum of the order of one tenth of collisional widths. This is not negligible, but clearly not sufficient to broaden significantly the main absorption features.

\begin{figure}[h]
\vspace{6mm}
\centering
\includegraphics[width=8cm]{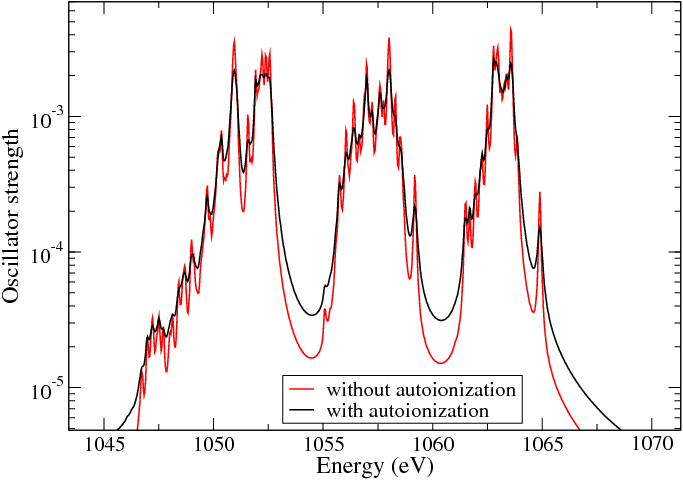}
\caption{Transition $1s^22s^12p^66d-1s^22s^12p^54d6d$ with and without broadening originating from autoionization process $1s^22s^12p^66d-1s^22s^22p^5\epsilon\ell$.}
\label{figure_1s22s12p66d_bis_2}
\end{figure}

Figure \ref{figure_1s22s12p66d_bis_2} represents transition $1s^22s^12p^66d-1s^22s^12p^54d6d$ with and without broadening due to autoionization process $1s^22s^12p^66d-1s^22s^22p^5\epsilon\ell$ computed using FAC code \cite{gu08}. The gap is slightly filled, but the initial configuration is a rather highly excited one, and its probability, and therefore its contribution to the spectrum, are not dominant. We checked that the effect of ionization/recombination is much smaller on the resonance transition $1s^22s^22p^6\rightarrow 1s^22s^22p^54d^1$.

\begin{figure}[h]
\vspace{5mm}
\centering
\includegraphics[width=12cm]{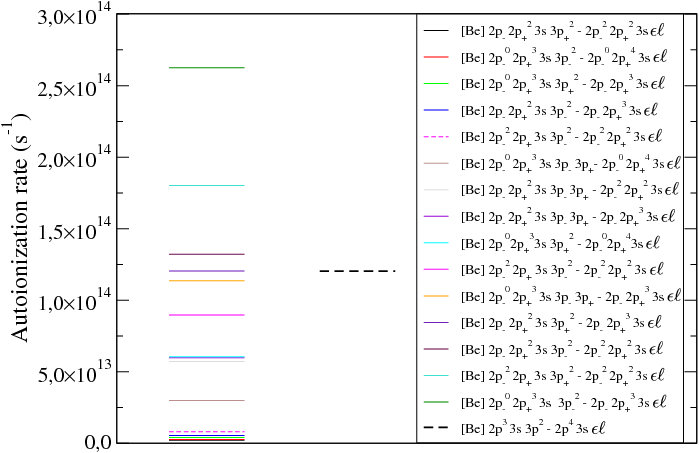}
\caption{Values of the autoionization rates between relativistic subconfigurations of $2p^33s^13p^2-2p^43s^1\epsilon\ell$. The dashed line represents the value between non-relativistic configurations $2p^33s^13p^2$ and $2p^43s^1\epsilon\ell$. Orbital $n\ell j$ is noted $n\ell_-$ if $j=\ell-1/2$ and $n\ell_+$ if $j=\ell+1/2$.}
\label{ato}
\end{figure}

We can see in Fig. \ref{ato} that the dispersion of autoionization rates between relativistic subconfigurations is likely to be rather important (see table \ref{tab_auto}).

\begin{table}[h]
\caption{Values of the configuration-averaged autoionization rate (s$^{-1}$) computed with FAC \cite{gu08} and MCDF \cite{bruneau83,bruneau84} codes for different channels.}
\centering
\begin{tabular}{ccc}\hline
%\toprule
\textbf{Autoionization process}	& \textbf{MCDF configurations} & \textbf{FAC configurations} \\\hline
%\midrule
$1s^22s^22p^43d^2$ – $1s^22s^22p^5\epsilon\ell$				& 7.146$\times$10$^{13}$ & 7.194$\times$10$^{13}$\\
$1s^22s^22p^43d^14d^1$ – $1s^22s^22p^5\epsilon\ell$	    	& 1.291$\times$10$^{13}$ & 1.374$\times$10$^{13}$\\
$1s^22s^22p^33d^2$ – $1s^22s^22p^4\epsilon\ell$				& 1.166$\times$10$^{14}$ & 1.221$\times$10$^{14}$\\
$1s^22s^22p^33p^14d^1$ – $1s^22s^22p^4\ell$					& 2.110$\times$10$^{13}$ & 2.214$\times$10$^{13}$\\
$1s^22s^22p^33s^13p^13d^1$ – $1s^22s^22p^43s^1\epsilon\ell$	& 1.894$\times$10$^{13}$ & 2.033$\times$10$^{13}$\\
$1s^22s^22p^43s^13p^14d^1$ – $1s^22s^22p^43p^1\epsilon\ell$	& 7.316$\times$10$^{11}$ & 6.621$\times$10$^{11}$\\
$1s^22s^22p^43p^2$ – $1s^22s^22p^5\epsilon\ell$				& 7.594$\times$10$^{13}$ & 8.156$\times$10$^{13}$\\
$1s^22s^22p^33s^13p^2$ – $1s^22s^22p^43s^1\epsilon\ell$		& 1.106$\times$10$^{14}$ & 1.204$\times$10$^{14}$\\
$1s^22s^22p^33s^13p^2$ – $1s^22s^22p^43p^1\epsilon\ell$		& 6.691$\times$10$^{13}$ & 7.484$\times$10$^{13}$\\\hline
%\bottomrule
\end{tabular}
\label{tab_auto}
\end{table}

\subsection{Breit interaction and QED corrections}

The Breit operator includes Coulomb repulsion, magnetic interaction and retardation in the electron-electron interaction due to finite value of the speed of light. The Breit Hamiltonian reads

\begin{equation}
h_B=\frac{1}{r_{12}}-\frac{\vec{\alpha}_1.\vec{\alpha}_2}{r_{12}}\cos\left(\omega_{12}r_{12}\right)+\left(\vec{\alpha}.\vec{\nabla}\right)_1\left(\vec{\alpha}.\vec{\nabla}\right)_2\frac{\cos\left(\omega_{12}r_{12}\right)-1}{\omega_{12}^2r_{12}},
\end{equation}

\noindent where $\alpha_i$ are the 4$\times$4 Dirac matrices, $\omega_{12}$ is the frequency of the exchange photon and the electron-electron interaction is expressed in the Coulomb (velocity) gauge \cite{grant74,grant07}.

\begin{figure}[h]
\vspace{2mm}
\centering
\includegraphics[width=90pt]{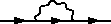}
\hspace{1cm}
\includegraphics[width=90pt]{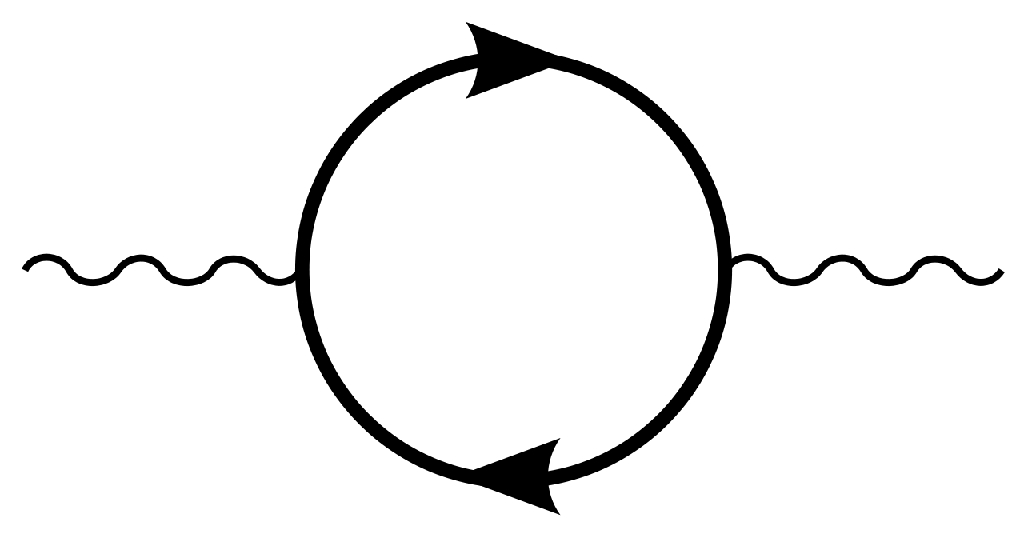}
\caption{Feynman diagrams for self-energy (left) and vacuum polarization (right).}
\label{fd}
\end{figure}

\begin{figure}[h]
\vspace{6mm}
\centering
\includegraphics[width=8cm]{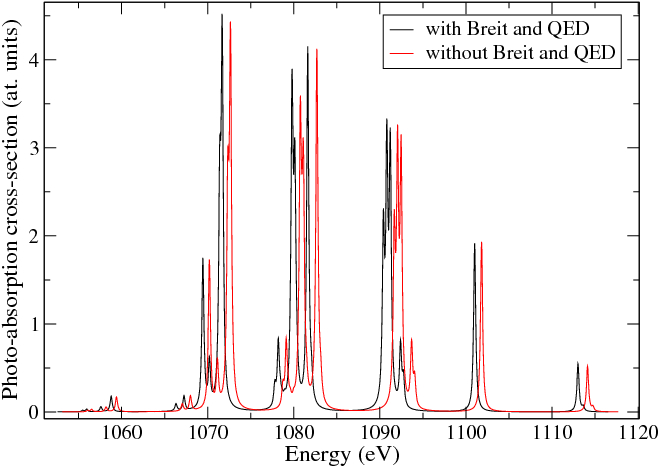}
\caption{Iron [Be] $2p^5 - 2p^44d$ structures of iron with and without Breit interaction and QED corrections. The calculations are perfomed with our MCDF code \cite{bruneau83,bruneau84}. SCO-RCG is quasi-relativistic and has no account for QED. Breit interaction is accounted for only partially.}
\label{mcdf_Fe_Breit_APS2016}
\end{figure}

In the present work, we consider two quantum electrodynamics (QED) corrections, responsible for the Lamb shift: vacuum polarization and self-energy (the corresponding Feynman diagrams are presented in Fig. \ref{fd}). Vacuum polarization is due to creation and annihilation of virtual electron-positron pairs in the field of the nucleus. The first term of order $\alpha(\alpha Z)$ is evaluated as the expectation value of the Uehling's potential \cite{uehling35,frolov12}. The self-energy (SE) represents the interaction of the electron with its own radiation field. In quantum field theory, this interaction corresponds to an electron emitting a virtual photon, which is then reabsorbed by the electron. In the early 1970s, Mohr proposed an atomic self-energy formulation \cite{mohr74} for hydrogenic atoms within the bound-state Furry formalism in a suitable form for numerical evaluation:

\begin{equation}\label{foncf}
E_{n\ell j}^{\mathrm{SE}}\left(\alpha Z\right)=\frac{\left(\alpha Z\right)^4}{\pi n^3\alpha}F_{n\ell j}\left(\alpha Z\right),
\end{equation}

\noindent where $F$ is a slowly varying function of $\alpha Z$. For $s$ and $p$ orbitals, $F$ is evaluated using a development in powers of $(Z\alpha)$ and $\ln(Z\alpha)$ for $Z\leq 10$ \cite{erickson77} and an interpolation in the tabulated values of Mohr \cite{mohr74,mohr75,mohr82} for $Z>10$. For $n$=3 and 4 we take the fit published by Curtis \cite{curtis85} and the results of Le Bigot \emph{et al.} \cite{lebigot01}. Calculation of many-electron radiative corrections is still one of the most difficult problems to deal with for high-precision level prediction. There have been no generalization of the self-energy calculations to arbitrary $N-$electron systems. Without exact solutions, atomic-structure codes use an approximation to the self-energy that consists of evaluating the exact hydrogenic formulas of Mohr and successors for an effective charge $Z_{\mathrm{eff}}$ in order to account for screening and multiple-electron interactions. The screening contribution to the self-energy is defined as

\begin{equation}
E_{n\ell j}^{\mathrm{SE}}\left(\alpha Z_{\mathrm{eff}}\right)-E_{n\ell j}^{\mathrm{SE}}\left(\alpha Z\right)=\frac{\alpha^3}{\pi n^3}\left(Z_{\mathrm{eff}}^4F_{n\ell j}\left(\alpha Z_{\mathrm{eff}}\right)-Z^4F_{n\ell j}\left(\alpha Z\right)\right),
\end{equation} 

\noindent where $Z_{\mathrm{eff}}$ takes into account screening by the other electrons. Such an approach is not as accurate as Welton's approach \cite{welton48}, but provides satisfactory results, when the effective charges are deduced from the average radius of the $n\ell j$ orbital \cite{pain17a}. We can see in Fig. \ref{mcdf_Fe_Breit_APS2016} that the shift due to Breit and QED correction is of the order of 2 eV. Because of the width of the experimental structures, it seems difficult to assert that such a shift brings a significant improvement of the computed line energies.

\section{Diagnostic utility of the 3C/3D line ratio in Fe XVII: another issue about iron...}\label{3c3}

Fe XVII X-ray emission is present in the coronae of the Sun, Capella ($\alpha$ Aurigae, located at 42.2 light-years from the Earth in the Auriga (or Cocher) constellation, third brightest star in the north hemisphere after Sirius and Arcturus) and Procyon. The latest, also known as $\alpha$ Canis Minoris, is the brightest star in the constellation of Canis Minor. It is a binary star, consisting of a white main-sequence star of spectral type F5 and a white dwarf companion named Procyon B, or NGC 4635. Two of the most distinct lines observed from Fe XVII are the resonance 3C $1s^22s^22p^53d$ $^1P_1$ $\rightarrow$ $1s^22s^22p^6$ $^1S_0$ and intercombination 3D $1s^22s^22p^53d$ $^3D_1$ $\rightarrow$ $1s^22s^22p^6$ $^1S_0$ lines. Values of the relative intensity $r$ of these two lines between 1.6 and 2.8 have been measured in non-flaring active regions of the solar corona. Values in the range 2.6 to 2.8 have been measured from Capella and a value of 1.8 has been reported for Procyon. However, there is a large discrepancy between observations and theory. Table \ref{3c3d} contains the values of the oscillator strengths of 3C and 3D lines calculated with our MCDF code \cite{bruneau83,bruneau84} in the ``transition state'' approximation within two different gauges: Coulomb (velocity) and Babushkin (length) \cite{grant74,grant07}. The self-consistent-field MCDF equations are obtained by requiring that a particular functional is stationary with respect to radial wavefunctions. Such a functional involves elements of the Hamiltonian matrix weighted by coefficients, usually referred to as ``generalized weights''. In the ``Average Level'' (AL) mode, the generalized weights are chosen to be the same for all the Atomic State Functions (ASF). Therefore, resolution of the self-consistent-field equations and diagonalization of the Hamiltonian matrix can be performed separately. However, if the initial state of a transition has a much larger degeneracy than the final state, the wavefunctions will describe preferentially the initial state (and \emph{vice versa}). In that case, orbital relaxation is not taken into account and transition energies may be inaccurate. The AL method can be improved using the Slater transition-state method \cite{slater74,bruneau83}, to equilibrate the weights between the initial and final configurations. 

In order to understand the discrepancy between experimental and theoretical values of $r$, a number of effects have been investigated, such as blending with an inner shell satellite line from Fe XVI, cascades from higher levels, non-Maxwellian electron distributions or configuration interaction \cite{fournier05}. The discrepancies between theory and these measurements prompted Bernitt \emph{et al.} \cite{bernitt12} to combine X-ray Free-Electron Laser (XFEL) at the Linac Coherent Light Source (LCLS) facility with an EBIT (Electron Beam Ion Trap). Well-defined X-ray pulses scanned the excitation energies of the associated states while simultaneously measuring the emission spectrum. The aim was to understand the discrepancies with the ratio measured in previous EBIT experiments. The LCLS experiment and the previous EBIT measurements are quite different in nature. The EBIT plasmas were electron-impact dominated, with many levels being simultaneously excited. The LCLS measurements were laser-driven and tuned to excite one transition at a time, producing a two-level system. The fluorescence intensity ratio of 3D and 3C was measured at $r$=2.61 $\pm$ 0.23, averaged over two independent measurement periods, after accounting for the blending of 3D with the Fe XVI line 3C. Loch \emph{et al.} \cite{loch15} performed a non-equilibrium modelling of the XFEL experiment and obtained a reduction of the predicted 3C/3D line ratio, but the discrepancy is still not fully understood. 

\begin{table}[h]
 \caption{Values of the oscillator strengths of 3C and 3D lines calculated with our MCDF code \cite{bruneau83,bruneau84} in the ``transition state'' approximation: $^1$ Coulomb (velocity) gauge, $^2$ Babushkin (length) gauge \cite{grant74,grant07}. The value of the ratio $r$ is not very sentitive to the choice of the gauge.}
\centering
\begin{tabular}{cccc}\hline
%\toprule
\textbf{Transition}	& \textbf{Energy (eV)} & \textbf{Oscillator strength$^1$} & \textbf{Oscillator strength$^2$}\\\hline
%\midrule
3C		& 811.86			& 1.820727$\times$10$^{-1}$	& 1.853965$\times$10$^{-1}$\\
3D		& 826.55			& 8.297194$\times$10$^{-1}$	& 8.432271$\times$10$^{-1}$\\\hline\hline
Ratio $r$ &                   & 4.557                 & 4.548\\\hline
%\bottomrule
\end{tabular}
\label{3c3d}
\end{table}

More generally, the emission arising from He-like iron lines and their dielectronic satellites, has been observed at low resolution in a number of clusters of galaxies. More precisely, using Gabriel's notation \cite{gabriel72}, the ratio $G=(x+y+z)/w$ of the intercombination ($x$ and $y$) and forbidden ($z$) to the resonance ($w$) lines arising from the $n$=2 level of the He-like ion, is a sensitive temperature diagnostics \cite{swartz93}.

\section{Stark effect, white dwarfs and Balmer lines}\label{whi}

\begin{figure}[h]
\vspace{2mm}
\centering
\includegraphics[width=6.2cm]{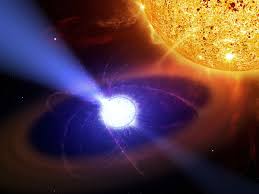}
\hspace{5mm}
\includegraphics[width=4.2cm]{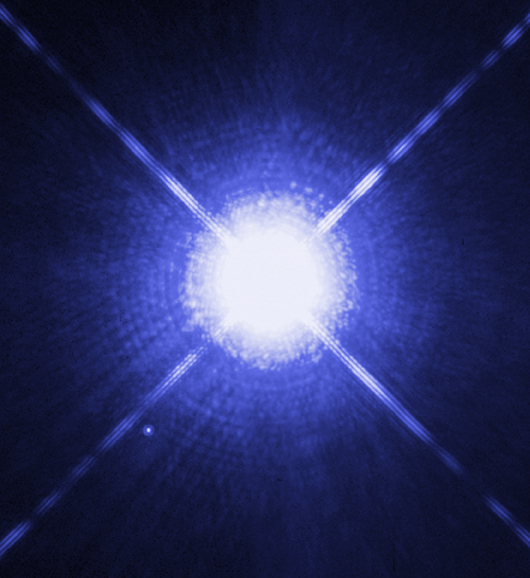}
\caption{Left: artist view of the cataclysmic variable binary star AE Aquarii, which consists of an ordinary star in close orbit around a magnetic white dwarf. \copyright Casey Reed (NASA). Right: Sirius A and its white dwarf companion star Sirius B (the small star on the left). \copyright NASA, ESA, H. Bond (STScI), and M. Barstow (University of Leicester).} 
\label{AquariiSirius}
\end{figure}

The diagnostic value of Stark-broadened lines such as H$_{\beta}$ ($n$=2 to $n$=4) has long been investigated by the experimentalists. It is well known that these lines play also a significant role in inferring the plasma conditions at the photospheres of high-gravity astronomical objects \cite{dimitrijevic15}. Particularly for the white dwarfs (WD), those gravities are $\approx$ 10$^4$ times higher than that of the Sun and their correspondingly higher electron densities ($n_e\approx$ 10$^{16}$-10$^{18}$ cm$^{-3}$) lead to significant Stark broadening of the lines. Exploiting this sensitivity, astronomers use the measured widths of these lines to infer the surface gravities of these WD and thereby determine their masses. Figure \ref{AquariiSirius} (left) represents an artist view of the cataclysmic variable binary star AE Aquarii, consisting of an ordinary star in close orbit around a magnetic white dwarf. The white dwarf has 63 \% of the Sun mass, but a radius of only about 1 \% of the Sun. The white dwarf in the AE Aquarii system is the first star of its type known to give off pulsar-like pulsations that are powered by its rotation and particle acceleration. It has the shortest known spin period of any white dwarf, completing a full revolution every 33.08 seconds. Figure \ref{AquariiSirius} (right) represents Sirius A and its white dwarf companion star Sirius B. The distance between Sirius A and Sirius B varies from 8.2 to 31.5 AU (1 AU=1.49597870$\times$10$^{8}$ km). Sirius, also know as the ``Dog star'', reflecting its prominence in its constellation, Canis Major, is the brightest star in the Earth's night sky.

The spectroscopic method is the most widely-used technique and is responsible for determining parameters for tens of thousands of WD. Obtaining high-accuracy estimates of those parameters is crucial for the use of WD, in order to determine the age of the universe, constrain the mass of supernovae progenitors and probe properties of dark matter axions.

Wiese \emph{et al.} measured the Balmer lines H$_{\alpha}$ , H$_{\beta}$ , H$_{\gamma}$ and H$_{\delta}$ in high-current, wall-stabilized arcs in hydrogen \cite{wiese72}. In 2009, Tremblay and Bergeron \cite{tremblay09} developed a new treatment of line broadening including non ideal effects, using the Hummer-Mihalas equation of state. They found a revision of 500-1000 K in temperature and 5-10 \% mass for the population of WD.

More recently Falcon \emph{et al.} \cite{falcon13,montgomery15} developed a platform on the Z machine at SNL to measure H$_{\beta}$ line shapes in photospheric conditions ($T\approx$ 1 eV, $n_e\approx$ 10$^{17}$ cm$^{-3}$), using the powerful X-ray capability of the machine to drive plasma formation in a gas cell. They measured hydrogen Balmer lines in emission and, for the first time, in absorption.

\begin{figure}[h]
\vspace{4mm}
\centering
\includegraphics[width=7.5cm]{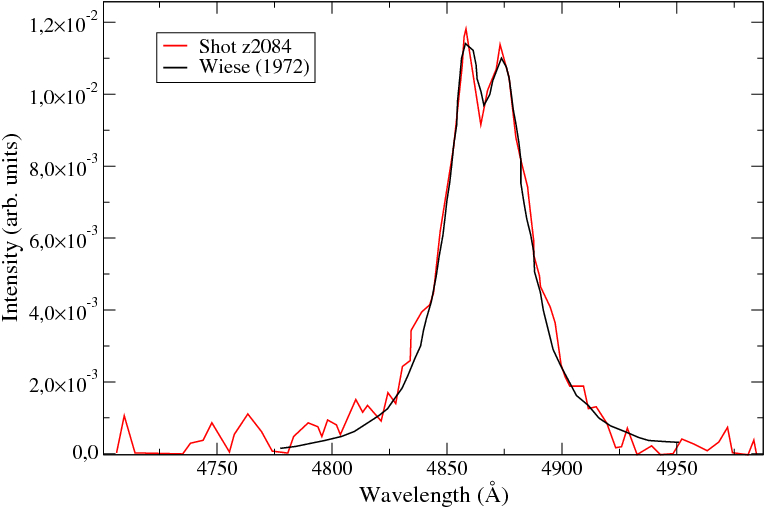}
\hspace{2mm}
\includegraphics[width=7.3cm]{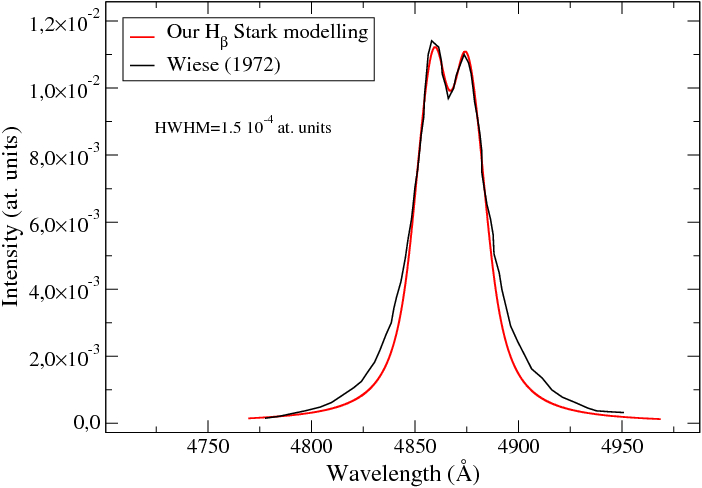}
\caption{Left: comparison between the H$_{\beta}$ profile measured by Wiese at $T$=1.15 eV and $n_e$=8.3$\times$10$^{16}$ cm$^{-3}$ and the experimental result obtained at SNL at $n_e$=5.76$\times$10$^{16}$ cm$^{-3}$ and proton density $n_H$=1.55$\times$10$^{17}$ cm$^{-3}$ (Shot z2084). Right: calculation with our line-shape code (the electric field is chosen equal to 2$\times$10$^{-5}$ atomic units and the profile is convolved with a Lorentzian of HWHM=1.5$\times$10$^{-4}$ atomic units. The spectrum of Wiese is also represented.}
\label{white_dwarf+hbeta}
\end{figure}

Up to now, SCO-RCG provided a precise Stark modeling only for hydrogen- and helium-like ions \cite{pain16}. As an example, the hydrogen $H_{\beta}$ profile computed by SCO-RCG at the conditions of Wiese \emph{et al.} is shown in Fig. \ref{white_dwarf+hbeta}, right. For more complex ions, a cruder approach is used \cite{porcherot11}. 

It is worth mentioning that the existence of white dwarfs with a carbon \cite{dufour07} and oxygen \cite{kepler09} atmospheres was reported. We can expect that several charge states are present (not only hydrogen-like oxygen). Therefore, it is important, in that case, to be able to compute accurate Stark profiles for ions with more than two electrons \cite{gilleron}.

\clearpage

\section{Conclusion}

The detailed opacity code SCO-RCG has a great potential for astrophysical applications. It is also able to provide accurate Rosseland mean opacities and we have shown the interpretation of several laser and Z-pinch spectroscopy experiments. Large changes in stellar opacities may make it necessary to consider new or overlooked physical phenomena absent in present opacity models. This stresses the importance of laboratory experiments to guide and validate theory.

Recent measurements of the iron photo-absorption relevant to the solar interior show significant discrepancies with theoretical results. The data, in the absence of unidentified systematic experimental errors, reveal surprising unobserved photo-absorption phenomena in plasmas. A conjecture to reconcile the discrepancies is that some transitions are missing in the calculations. Another possibility is a redistribution of oscillator strengths not included in the models. A full configuration-interaction treatment, not performed in SCO-RCG calculations, may transfer oscillator strength to higher energies. As mentioned by Iglesias \cite{iglesias15b} the observed enhanced absorption seems incompatible with theory since calculations satisfy the Thomas-Reiche-Kuhn sum rule and preserve oscillator strength density and particle number. In this work, the disagreement between experiment and theory was examined, and although none of our investigations provide an explanation (density effects, highly excited states, line broadening, ...), this enigmatic spectrum remains definitely an exciting mystery and stimulates many new developments in opacity modelling and computation.

\vspace{5mm}

{\bf Acknowledgments}
The authors would like to thank all the authors of Refs. \cite{davidson88,winhart95,winhart96,dozieres15,bailey03,bailey07,dervieux15} for the experimental spectra as well as J. Colgan for providing the ATOMIC spectra.

\end{document}